\documentclass[sigconf, nonacm, screen]{acmart}

\usepackage{arydshln}

\usepackage{enumitem}
\usepackage{subcaption}
\usepackage{graphicx}
\usepackage[bottom]{footmisc}
\usepackage{xcolor}

\newtheorem{theorem}{Theorem}[section]

\newtheorem{remark}[theorem]{Remark}

\begin{document}

\title[Decoupling Inference and State Updates]{Decoupling Inference from State Updates in Low-Latency Feature Engines via Probabilistic Thinning}


\author{Augusto Peres}
\affiliation{%
  \institution{Feedzai}
  \city{Lisbon}
  \country{Portugal}
}
\email{augusto.peres@feedzai.com}

\author{Iker Perez}
\orcid{0000-0001-9400-4229}
\affiliation{%
  \institution{Feedzai}
  \city{London}
  \country{UK}
}
\email{iker.perez@feedzai.com}

\author{Pedro Valdeira}
\affiliation{%
  \institution{Feedzai}
  \city{Lisbon}
  \country{Portugal}
}
\email{pedro.valdeira@feedzai.com}

\author{Guilherme Jardim}
\affiliation{%
  \institution{Feedzai}
  \city{Lisbon}
  \country{Portugal}
}
\email{guilherme.jardim@feedzai.com}

\author{Ana Sofia Gomes}
\affiliation{%
  \institution{Feedzai}
  \city{Lisbon}
  \country{Portugal}
}
\email{sofia.gomes@feedzai.com}

\author{Hugo Ferreira}
\affiliation{%
  \institution{Feedzai}
  \city{Lisbon}
  \country{Portugal}
}
\email{hugo.ferreira@feedzai.com}

\author{Pedro Bizarro}
\affiliation{%
  \institution{Feedzai}
  \city{Lisbon}
  \country{Portugal}
}
\email{pedro.bizarro@feedzai.com}

\renewcommand{\shortauthors}{Peres et al.}

\begin{abstract}
Streaming data systems increasingly underpin Machine Learning workflows that maintain large numbers of continuously updated aggregations. In production settings, each incoming event typically triggers read--modify--write operations to persistent storage, making high-frequency state updates a dominant source of latency, contention, and operational cost. 
In this work, we decouple inference from state persistence in streaming Machine Learning pipelines via probabilistic thinning: every event is scored, but durable state updates are selectively triggered by \textit{informative events}. Unlike approaches that shed input or state, we show that persistence-path control is achievable without a high-frequency in-memory control plane or cross-worker coordination, relying exclusively on approximate statistics retrieved from disk-backed key-value stores. 
We model the resulting stochastic processes, derive bounds on filtering rates, and prove that common time-based aggregations remain unbiased under variance-aware formulations, preventing systemic error accumulation. We evaluate the approach in a controlled setting that isolates per-event costs, demonstrating substantial reductions in storage Input/Output and serialization overhead. Across experiments, up to 90\% of events are excluded from the persistence path while preserving and in some cases improving downstream utility.
\end{abstract}





\maketitle

\section{Introduction}
\label{sec:introduction}

Streaming \textit{Machine Learning} (ML) workflows underpin a wide range of critical applications, including fraud detection, recommender systems, and cybersecurity. In these settings, decisions must be made continuously over high-volume event streams under strict latency constraints. A common architectural pattern is to maintain \emph{features} or \emph{profiles}: temporal aggregations such as counts, sums, or averages computed per uniquely identifiable \emph{entities} in the data, including payment cards, users, or IP addresses~\cite{wong2024training, whitrow2009transaction, dal2014learned, le2004machine, verwiebe2023survey}.

\begin{figure*}[h] 
    \centering
        \begin{minipage}{1.0\textwidth}
            \centering
            
            \begin{subfigure}{0.48\textwidth}
                \centering
                \includegraphics[width=\linewidth]{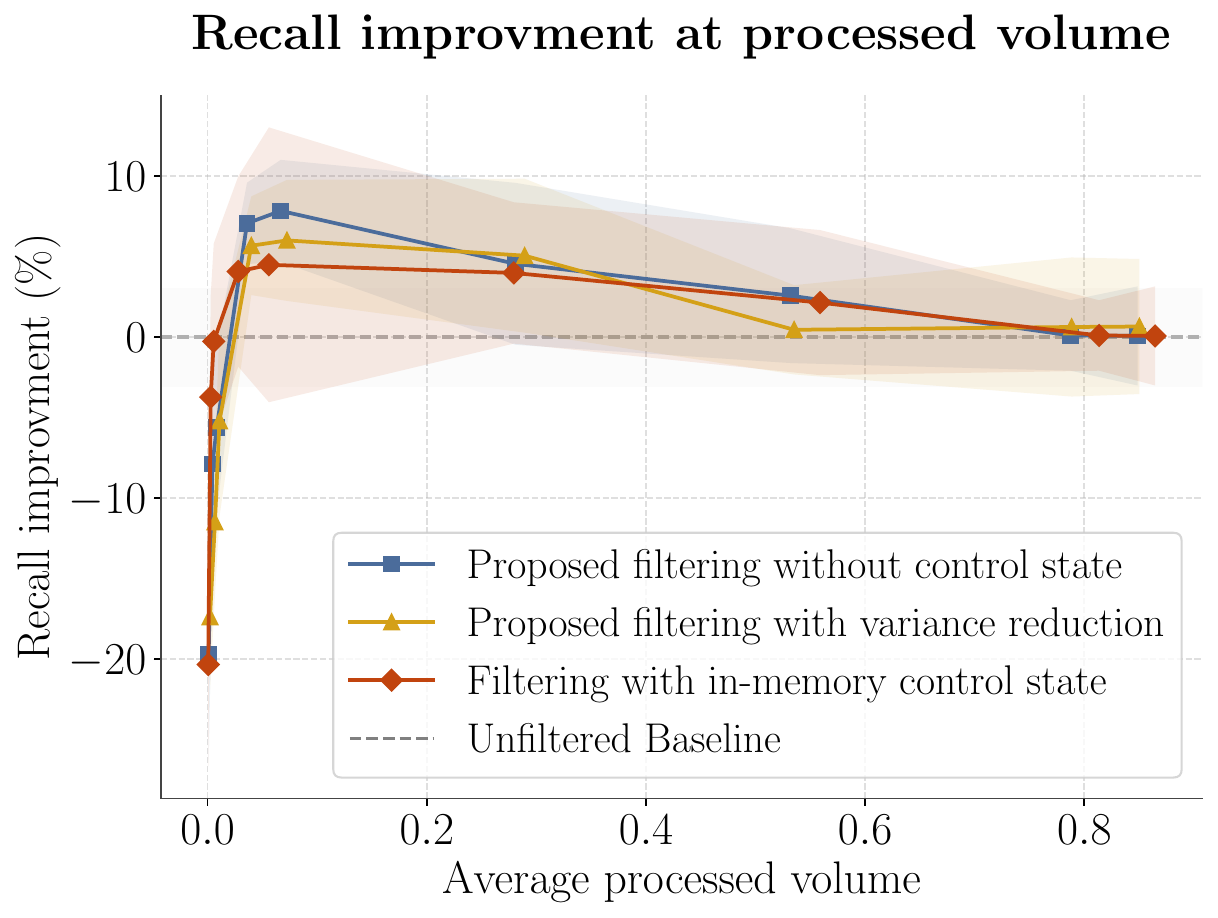}
            \end{subfigure}
            \hfill
            \begin{subfigure}{0.48\textwidth}
                \centering
                \includegraphics[width=\linewidth]{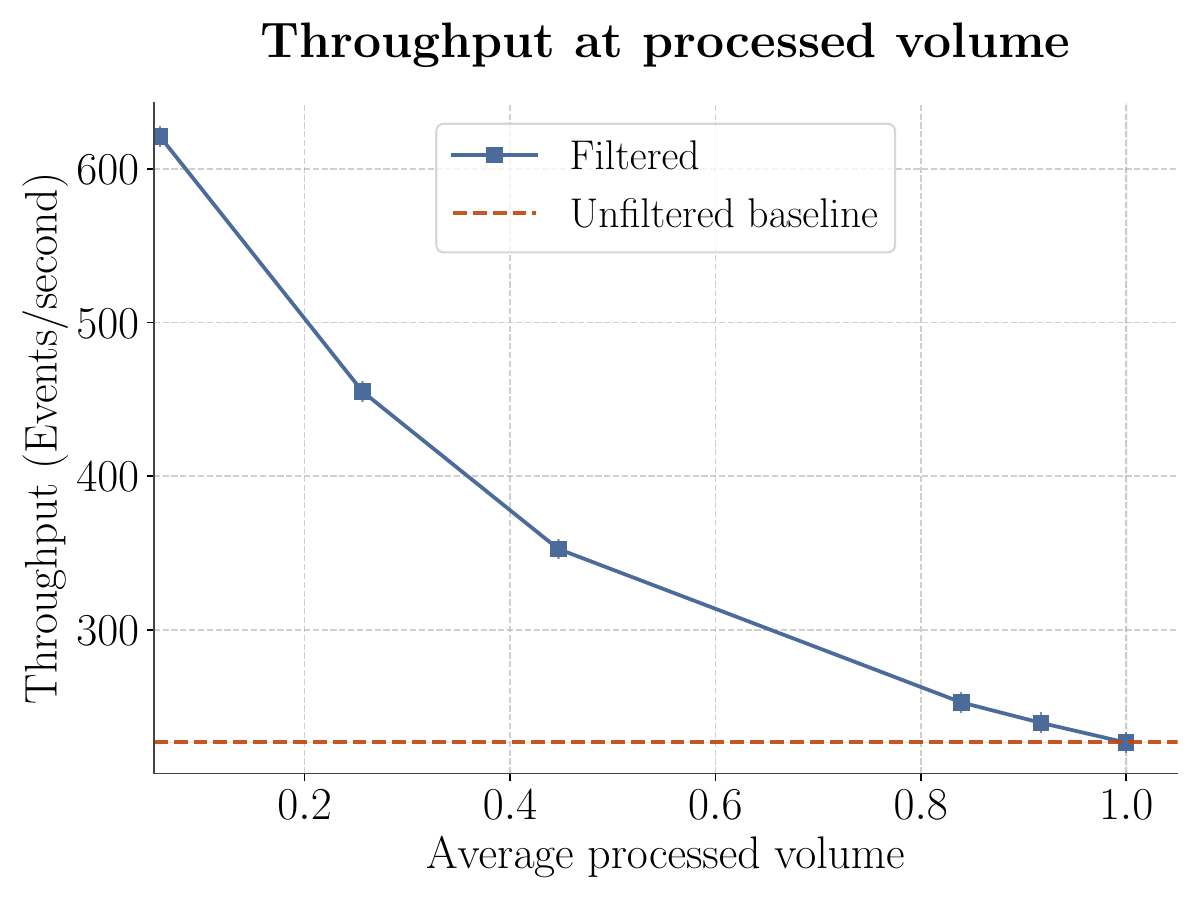}
            \end{subfigure}
            
            \vspace{-1em}
        \end{minipage}
    \caption{Performance benefits of persistence-path control on real-world data. Left: ML utility vs. filtered data volume for a binary classification task. Selectively discarding events from persistence path operations, but still scoring every event, preserves or improves predictive performance. Right: System throughput vs. persistence-update frequency. Reducing the fraction of events that trigger costly state updates enables a substantial increase in end-to-end system capacity.}
    \label{fig:teaserMain}
\end{figure*}

However, maintaining feature profiles at scale incurs substantial operational costs. In modern streaming architectures, each incoming event typically triggers a \textit{read--modify--write} (RMW) interaction with persistent state: existing aggregates are retrieved, updated, and written back to storage. As event rates increase, the tight coupling between event processing and state persistence becomes a dominant source of latency, contention, and infrastructure cost~\cite{jang2020specialized}.

To manage costs, modern production systems rely on a combination of distributed stream processing and approximate computation. Streaming platforms such as Kafka Streams~\cite{kreps2011kafka}, Apache Flink~\cite{carbone2017state}, or Spark Structured Streaming~\cite{armbrust2018structured} partition event streams across workers and maintain per-entity state locally. At the same time, approximation techniques, including sketches, recursive estimators, or sampling, are widely used to reduce memory footprint or per-update computation~\cite{approximate-computing, synopses-for-massive-data, CORMODE200558}. While effective, these approaches largely treat state updates as unavoidable side effects of event processing. In domains where discarding or deferring events is unacceptable, such as fraud detection, every event triggers a persistence update, leaving state maintenance as a fundamental scalability bottleneck.

In this work, we show that this limitation is not fundamental. A key observation is that streaming ML workloads commonly exhibit pronounced \emph{data skew}: a small fraction of entities account for a disproportionate share of events. While this is traditionally viewed as a systems challenge, we show that it enables selective suppression of persistence operations without sacrificing inference quality. We introduce \emph{persistence-path control}, a probabilistic thinning and state management mechanism that decouples inference from persistence. Every incoming event is evaluated by the ML model, but only a subset triggers updates to feature profiles. In contrast to classical thinning or load-shedding based on external control or input-driven policies, inclusion decisions are driven exclusively by disk-backed approximate statistics, governed by a self-correcting stochastic process that preserves unbiasedness and provides explicit variance control. While prior work explores utility-driven shedding over intermediate state, we instead enable \textit{stateless orchestration} embedded in the persistence path, eliminating high-frequency in-memory counters, coordination, or auxiliary control.

The design targets a broad class of stateful, decayed and recursive per-entity aggregations, while preserving unbiased estimates, enabling variance reduction, and dramatically reducing the frequency of expensive write operations. Figure~\ref{fig:teaserMain} previews the resulting trade-offs: large reductions in storage \textit{Input/Output} (I/O) and serialization overhead translate into higher throughput and, remarkably, improved downstream ML performance due to an implicit regularization effect under highly skewed workloads.

\paragraph{Statement of Contributions.}
This paper demonstrates how probabilistic thinning can be designed and operationalized to deliver scalable streaming ML systems. Our main contributions are:
\begin{itemize}
    \item We identify the tight coupling between event processing and persistent state updates as a primary scalability bottleneck in modern streaming ML architectures; see Section~\ref{sec:adaptive-event-filtering}.
    \item We formulate persistence-path control as a stateless orchestration mechanism for thinning without local state or cross-worker coordination; see Section~\ref{sec:compute-constrains}.
    \item We model the resulting process and derive analytical bounds on filtering rates, showing unbiasedness and variance control for common aggregations; see Section~\ref{sec:compute-constrains}.
\end{itemize}
Finally, in Sections~\ref{sec:system-implementation} and ~\ref{sec:evaluation} we discuss operating assumptions, and evaluate the approach in controlled settings that isolate per-event persistence costs and avoid confounding runtime effects. Our results demonstrate reductions of up to $90\%$ in storage I/O, serialization and replication traffic, as well as $45\%$ reduction in overall system utilization, while preserving and in several cases improving downstream ML utility across domains.

\section{Background}
\label{sec:background}

Streaming ML systems are designed to make predictions or decisions over continuous event flows under strict latency and reliability constraints. In domains such as fraud detection (a recurring example throughout this work), recommender systems, and cybersecurity, \textit{Service Level Agreements} routinely require millisecond-level response times at extreme percentiles (e.g., 250\,ms at the 99.9th) under sustained high throughput~\cite{railgun,stream,ramasamy}. Meeting these requirements at scale places streaming ML at the intersection of distributed systems engineering and approximate computation.

\subsection{State Management in Stream Processing}

To sustain workloads processing millions of events per second, modern platforms such as Kafka Streams~\cite{kreps2011kafka}, Apache Flink~\cite{carbone2017state}, Spark Structured Streaming~\cite{armbrust2018structured}, and related systems~\cite{railgun,stream,samza,ramasamy} rely on distributed architectures that partition workloads across independent workers. Each worker maintains persistent local state in embedded \textit{key--value} (KV) stores, commonly backed by \textit{Log-Structured Merge Trees} (LSM-Trees)~\cite{ONeil1996} (e.g., RocksDB~\cite{pamarthi2024real}). For every incoming event, a worker must \textit{deserialize} historical state, apply an update, and \textit{serialize} the result back to storage.

While this architecture enables horizontal scalability and minimizes coordination, it incurs substantial overheads associated with the RMW execution model:
\begin{itemize}
    \item \emph{State amplification:} LSM-Trees suffer from write amplification during compactions and read amplification from multi-level lookups and Bloom filter checks~\cite{ONeil1996, dong2021rocksdb}.
    \item \emph{Serialization overhead:} In JVM-based systems~\cite{javavm}, serializing state into bytes for storage and deserializing it for updates often dominates CPU time, creating a pronounced \textit{serialization--deserialization} (SerDe) bottleneck~\cite{jang2020specialized}.
\end{itemize}
These costs compound for high-frequency keys, where even lightweight updates translate into sustained I/O pressure and elevated tail latency.

\subsection{Mathematical Foundations for Efficiency}

Approximate computation has long been employed to reduce the cost of maintaining state in large-scale data systems~\cite{approximate-computing,synopses-for-massive-data}. Classical work in approximate query processing~\cite{synopses-for-massive-data} and stochastic simulation~\cite{kroese2013handbook} showed how controlled uncertainty can yield significant efficiency gains. In streaming settings, exponential decay mechanisms~\cite{maintain-time-decayed-agg-over-streams,verwiebe2023survey}, incremental estimators such as Welford’s method~\cite{Welford01081962}, and related recursive formulations~\cite{brown1956exponential,Hunter01101986,a4968c25c2844c1b9c1ed4f3e0d45b01} reduce aggregation updates to constant-time operations, enabling long event histories to be summarized in bounded memory. 

Probabilistic data structures, including Bloom filters~\cite{bloom-filters-and-applications}, count--min sketches~\cite{CORMODE200558}, and HyperLogLog~\cite{flajolet2007hyperloglog}, provide sublinear-memory summaries with bounded bias and variance. While these techniques optimize the \emph{cost per updates}, even compact sketches must be retrieved from storage, deserialized, updated, and written back. For high-velocity entities, such as popular merchants in payment streams, the dominant bottleneck is therefore not storage capacity, but the sustained rate of persistence operations; such as I/O \textit{Operations Per Second} (IOPS) and CPU SerDe processes. Our work targets the complementary dimension: reducing the \emph{frequency} of state updates while preserving statistical correctness under controlled stochastic selection.

\subsection{Load Shedding and Stream Sampling}

When input rates exceed capacity, streaming engines commonly resort to load shedding or sampling. Ingestion-level techniques such as reservoir sampling or threshold-based shedding discard events early to protect downstream operators~\cite{Abadi2005TheDO, babcock2002models}, while selective approaches prioritize events based on utility or query relevance~\cite{tatbul2006window, quoc2017streamapprox}. Alternatively, state-shedding methods leverage in-memory buffers to dynamically select events under pressure~\cite{zhao2020load, slo2020state}, trading off inference completeness by evicting intermediate state under windowed, query-driven semantics.

Although effective, shedding is incompatible with security and risk-sensitive ML applications. In fraud detection, for example, every transaction must be inspected: discarding an events at ingestion or evicting intermediate state compromises both inference coverage and long-lived feature consistency. Furthermore, reliance on complex in-memory control logic or cross-worker coordination to manage such shedding can become a scalability bottleneck. Our work builds on these foundations but instead applies stochastic thinning directly at the persistence layer. We treat state updates as probabilistic events decoupled from inference triggers, enabling persistence-path control that avoids both input dropping and state eviction, while eliminating high-frequency in-memory control. Thus, we reduce write frequency and mitigate the RMW and SerDe bottlenecks intrinsic to modern streaming engines.

\section{Adaptive Event Filtering}
\label{sec:adaptive-event-filtering}

\begin{figure}[t]
    \centering
    \includegraphics[width=\columnwidth]{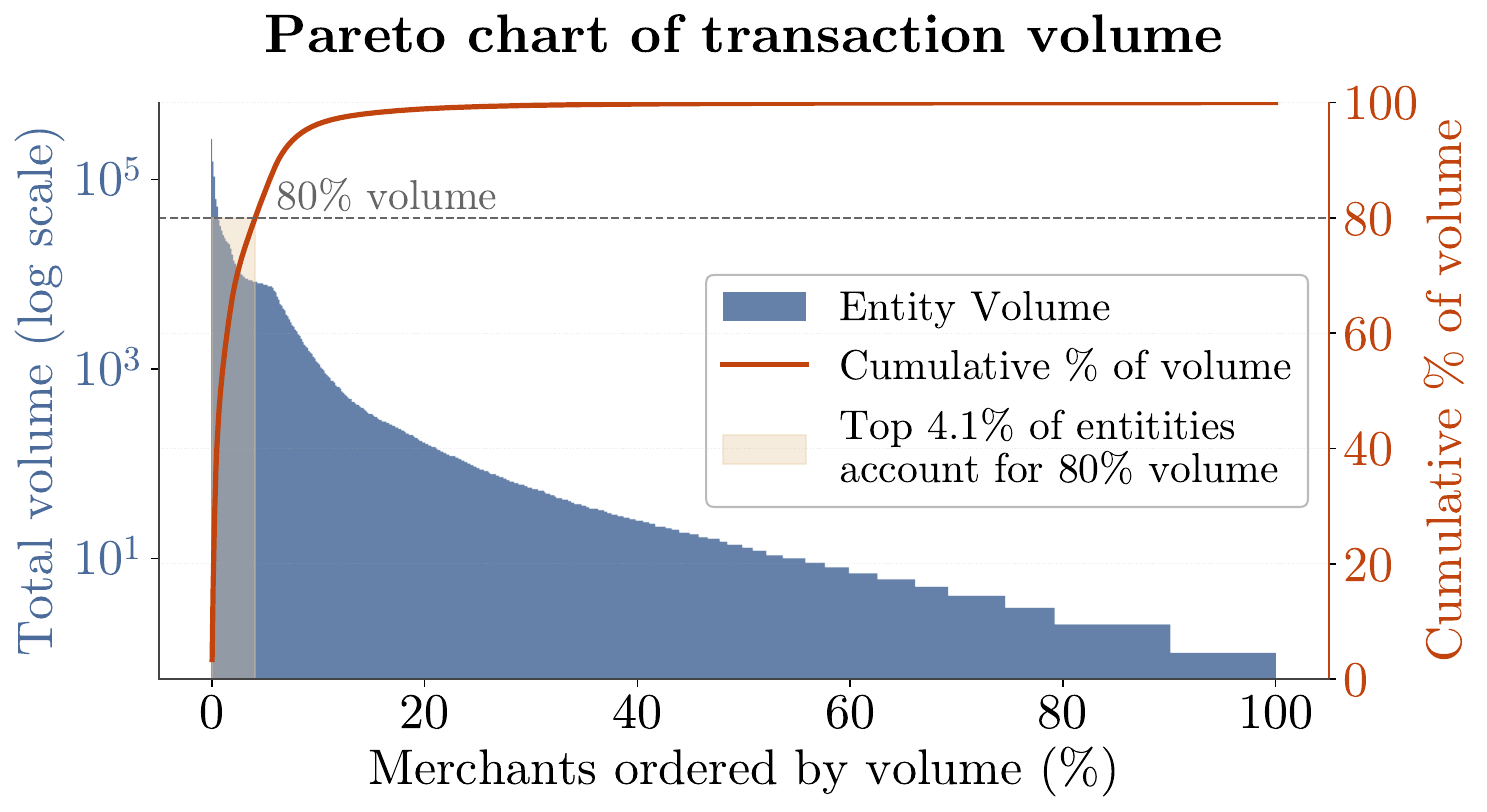}
    \caption{Merchant activity volume in a real-world transaction dataset, over a 7 month interval. Transaction counts (log scale) are shown for unique merchants ordered by volume, revealing a highly skewed distribution.}
    \label{fig:entity-histograms}
\end{figure}

Streaming ML workloads exhibit strong heterogeneity in event arrival patterns across entities. In financial transaction monitoring, activity is often concentrated on a small subset of accounts, while most entities remain inactive for extended periods. Figure~\ref{fig:entity-histograms} illustrates this effect on real-world data: approximately 4\% of merchant entities generate nearly 80\% of incoming events.

\begin{figure*}[h!]
    \centering
    \includegraphics[width=1.0\textwidth]{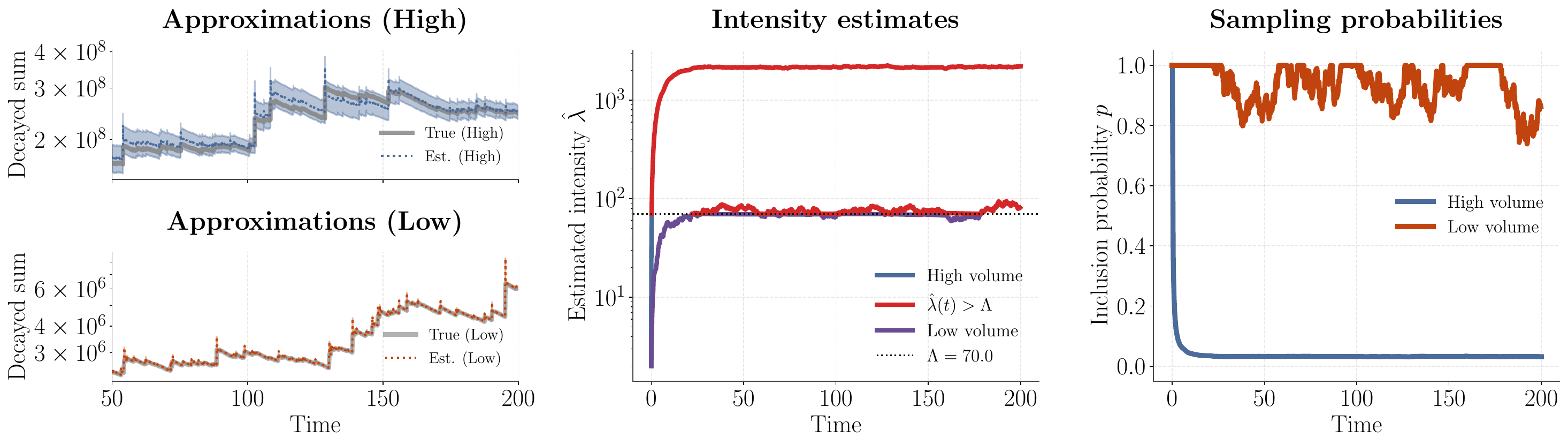}
    \caption{Adaptive event filtering applied to high- and low-velocity entities. Left: Reconstructed decayed sums for LogNormal distributed $\{q_n\}_{n=1,2,\dots}$, averaged over $100$ runs and cropped to $t\in[50,200)$ for readability. Center: Estimated arrival rates $\hat{\lambda}(t)$ compared to a thinning \textit{threshold} $\Lambda$. Right: Probability of triggering persistence-path write operations over time. High-intensity entities undergo aggressive filtering (fewer writes), while low-intensity entities remain largely unaffected.}
    \label{fig:event-thinning}
\end{figure*}

Under such skew, executing a full RMW cycle for every event yields diminishing returns. For high-volume entities, successive updates contribute little new statistical information while incurring substantial serialization, I/O, and replication cost. This motivates \emph{adaptive event filtering}: a systems-level mechanism that suppresses persistent state updates while preserving inference on every event, under a controlled probabilistic thinning policy~\cite{kingman1992poisson, daley2003introduction}. 

In this work, we frame this as \emph{intensity-aware thinning} of stochastic processes~\cite{kingman1992poisson,daley2003introduction}, adapting ideas from load shedding~\cite{tatbul2003load}, rate limiting~\cite{cardellini2022runtime}, or adaptive sampling~\cite{synopses-for-massive-data,duffield2017stream,ting2022adaptive}, but shifting their purpose. Rather than controlling ingestion, we bound write frequency and serialization cost under skewed ML workloads, while maintaining unbiased aggregates. We first describe the process under the standard assumption that control statistics are continuously updated in-memory. Later, we address the significant challenges of maintaining these statistics in modern streaming systems.

\subsection{Problem Formulation}  
For simplicity and without loss of generality, we consider a time-ordered event stream
\[
    \mathcal{E} = \{ e_1, e_2, \dots \}, \quad \textup{with} \quad \mathcal{E}(t) = \{e_n\in \mathcal{E}: t_n \leq t \},
\]
associated with a single entity, such as a merchant, card number, or device. Each event $e_n = (q_n, t_n)$ consists of a quantitative attribute $q_n \in \mathbb{R}$ (e.g. transaction amount) and a timestamp $t_n \in \mathbb{R}^+$.

The arrivals form a marked inhomogeneous Poisson Point Process~\cite{daley2003introduction, jacobsen2006point} with \textit{intensity} function \(\lambda(t)\). Thus, the expected number of events $N$ in any interval $(t_1, t_2]$ is given by:
\[
    \mathbb{E}\!\left[N(t_1,t_2)\right] = \int_{t_1}^{t_2} \lambda(t)\,dt,
\]
In practice, \(\lambda(t)\) varies widely across entities and over time, reflecting behavioural heterogeneity. Figure~\ref{fig:event-thinning} illustrates this effect for two entities with high and low arrival intensity.
 
\begin{figure}[h]
    \centering
    \includegraphics[width=0.98\columnwidth]{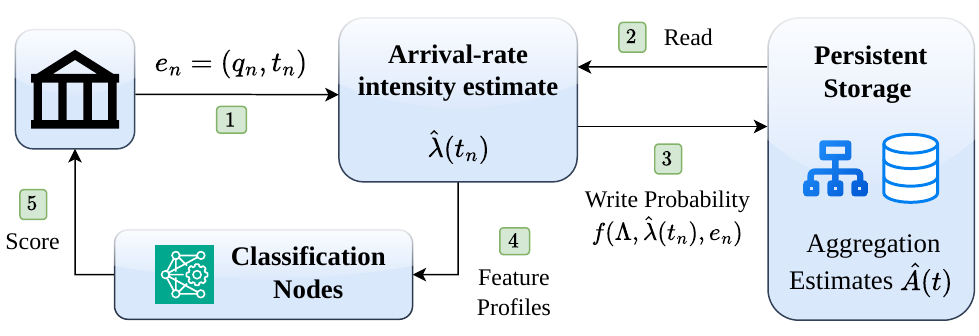}
    \caption{Naive adaptive filtering. (1) Event arrives; (2) System retrieves aggregations from persistent storage; (3) Computes updates and writes back to persistence \textit{probabilistically}; (4) Event is sent for scoring and (5) returned to the client.}
    \label{fig:probabilistic-system}
\end{figure}

While every event $e_n$ arriving at the system triggers an inference task, we execute a persistent state update \textbf{only if} an auxiliary Bernoulli variable \(Z_n \sim \mathcal{B}(p_n)\) evaluates to $1$. This yields a filtered sub-stream of persisted events:
\[
    \mathcal{I} = \{e_n \in \mathcal{E}: Z_n = 1\}, \quad \text{with} \quad \mathcal{I}(t) = \{e_n\in \mathcal{I}: t_n \leq t \}.
\]
The \textit{inclusion probability} \(p_n\) governs the expected write rate, and is determined by a policy
\begin{equation}
    p_n = f(\Lambda, \hat{\lambda}(t_n), e_n), \label{eq:downsampling-prob}
\end{equation}
which enforces a \textit{user-defined} write budget \(\Lambda\) relative to an estimate \(\hat{\lambda}(t_n)\) of the \textit{unknown} true event intensity\footnote{Processes to estimate \(\lambda(t)\) from observed event histories are detailed in Section~\ref{sec:compute-constrains}.}. A common baseline is
\begin{align}
    f(\Lambda, \hat{\lambda}(t_n), e_n) = \min\left(1, \frac{\Lambda}{\hat{\lambda}(t_n)}\right),
    \label{eq:naive}
\end{align}
which guarantees \(\mathbb{E}[\sum_{i=1}^n Z_i] \le \Lambda \cdot t_n\) and throttles persistence-path operations for high-velocity entities. Figure~\ref{fig:probabilistic-system} illustrates a schematic of this \textit{naive} workflow.

\subsection{Probabilistic State Management}

\begin{figure*}[t]
    \centering
    \includegraphics[width=0.98\textwidth]{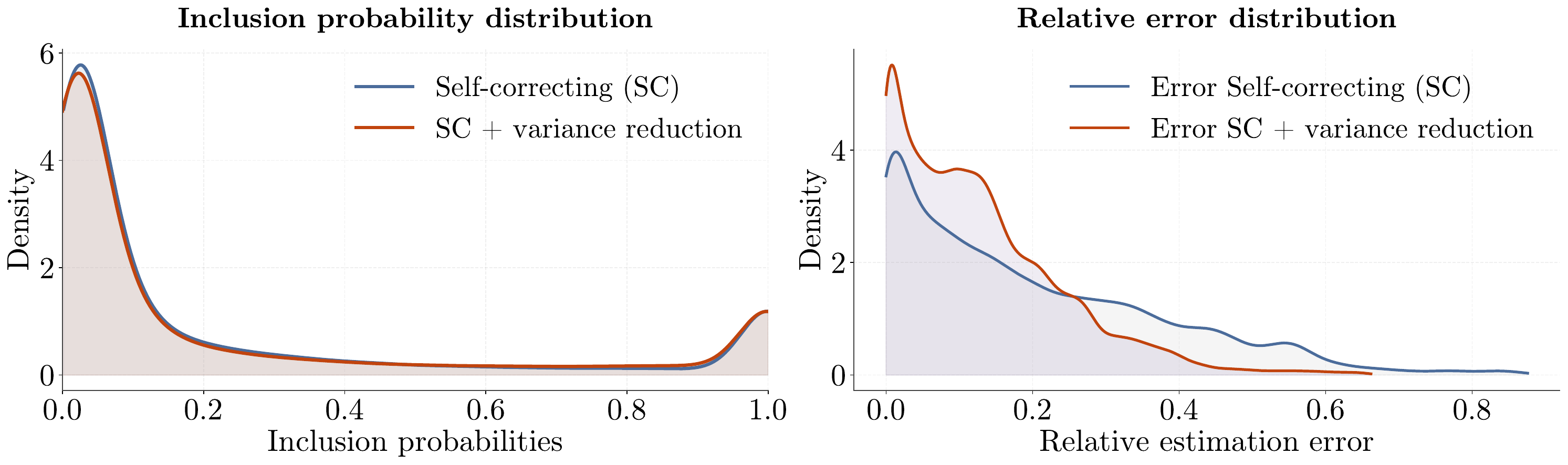}
    \caption{Inclusion probabilities and estimator error across \textit{sum}, \textit{count}, and \textit{average} aggregations $A(t)$. We compare standard (blue) and variance-reduced (red) filtering strategies. Left: Variance-aware strategies reallocate probability mass for write cycles while maintaining total write budget. Right: Targeted reallocation reduces estimation error without increasing system load.}
    \label{fig:inclusion-probability-dist}
\end{figure*}

Let \(A(t)\) denote an aggregate computed over the full stream, such as a \textit{sum} or \textit{count}, with per-event contribution \(w(t, e_n)\). Under adaptive filtering, the system introduces stochastic sparsity by suppressing updates. We therefore approximate \(A(t)\) using a Horvitz-Thompson estimator~\cite{Horvitz1952AGO} over the sub-stream $\mathcal{I}(t)$, i.e.
\[
    \widehat{A}(t) = \sum_{e_n \in \mathcal{I}(t)} \frac{w(t,e_n)}{p_n}.
\]
This estimator inversely scales contributions by their inclusion probability, remains unbiased, and has variance:
\begin{align}
    \label{eq:variance-estimators}
    \text{Var}\left[\widehat{A}(t)\right] = \sum_{e_n \in \mathcal{E}(t)} w^2(t,e_n)\left(\mathbb{E}\left[\frac{1}{p_n}\right] - 1\right).
\end{align}
Thus, approximation error is dominated by large-magnitude contributions \(w(t,e_n)\) retained with low probability. We refer the reader to Appendix~\ref{app:estimator-proofs} for the derivation through conditioning on $\hat{\lambda}(t_n)$.

\subsection{Recursive Aggregates}

Streaming feature stores typically maintain aggregates over complete histories, using exponentially decayed counts, sums, or averages, because they admit constant-space updates compatible with KV-based state stores~\cite{cormode2008exponentially, maintain-time-decayed-agg-over-streams, roberts2000control}. Table~\ref{tab:core_metrics} outlines standard per-event weights $w(t, e_n)$, where $\tau > 0$ represents a \textit{decay factor} analogous to sliding-window intervals.

\begin{table}[h]
    \centering
    \caption{Examples of popular per-event contribution functions $w(t, e_n)$ used in streaming feature stores.}
    \label{tab:core_metrics}
    \begin{tabular*}{\columnwidth}{@{\extracolsep{\fill}}lcc@{\extracolsep{\fill}}}
    \toprule
    \multicolumn{1}{c}{\textbf{Metric Type}} & \multicolumn{1}{c}{\textbf{Count Weight}} & \multicolumn{1}{c}{\textbf{Sum Weight}}\\
    \cmidrule(lr){2-2} \cmidrule(lr){3-3} 
    \textbf{All-Time} & $1$ & $q_n$ \\
    \textbf{Exponential Decay} & $\exp\left(-\frac{|t-t_n|}{\tau}\right)$ & $q_n \cdot \exp\left(-\frac{|t-t_n|}{\tau}\right)$ \\
    \bottomrule
    \end{tabular*}
\end{table}

When enforcing thinning, update rules for recursive aggregates elegantly incorporate a Bernoulli mask, s.t.
\[
    \widehat{A}(t_n) = Z_n \cdot \frac{w(t_n, e_n)}{p_n} + \exp\left(-\frac{t_n-t_{n-1}}{\tau}\right) \widehat{A}(t_{n-1}).
\]
This simple formulation supports, without loss of generality, for the measurement of \textit{averages}, \textit{ratios}, \textit{squared means}, \textit{moments}, or \textit{variances}, to name only a few. 

\subsection{Variance-Aware Adaptive Filtering}

For an unbiased estimator \(\widehat{A}\) of the true aggregate \(A\), precision is often measured by the coefficient of variation~\cite{everitt2010cambridge}; i.e.
\[
    \mathrm{CV}\left(\widehat{A}(t)\right) = 
    \frac{\sqrt{\text{Var}[\widehat{A}(t)]}}{\mathbb{E}[\widehat{A}(t)]} 
    \propto \frac{1}{\sqrt{|\mathcal{I}(t)|}},
\]
where \(|\mathcal{I}(t)|\) is the number of retained events. As volume grows, the marginal utility of each update for precision decreases. Adaptive filtering exploits this by suppressing updates beyond a threshold, in order to achieve system-level savings.

\begin{figure}[b]
    \centering
    \includegraphics[width=\columnwidth]{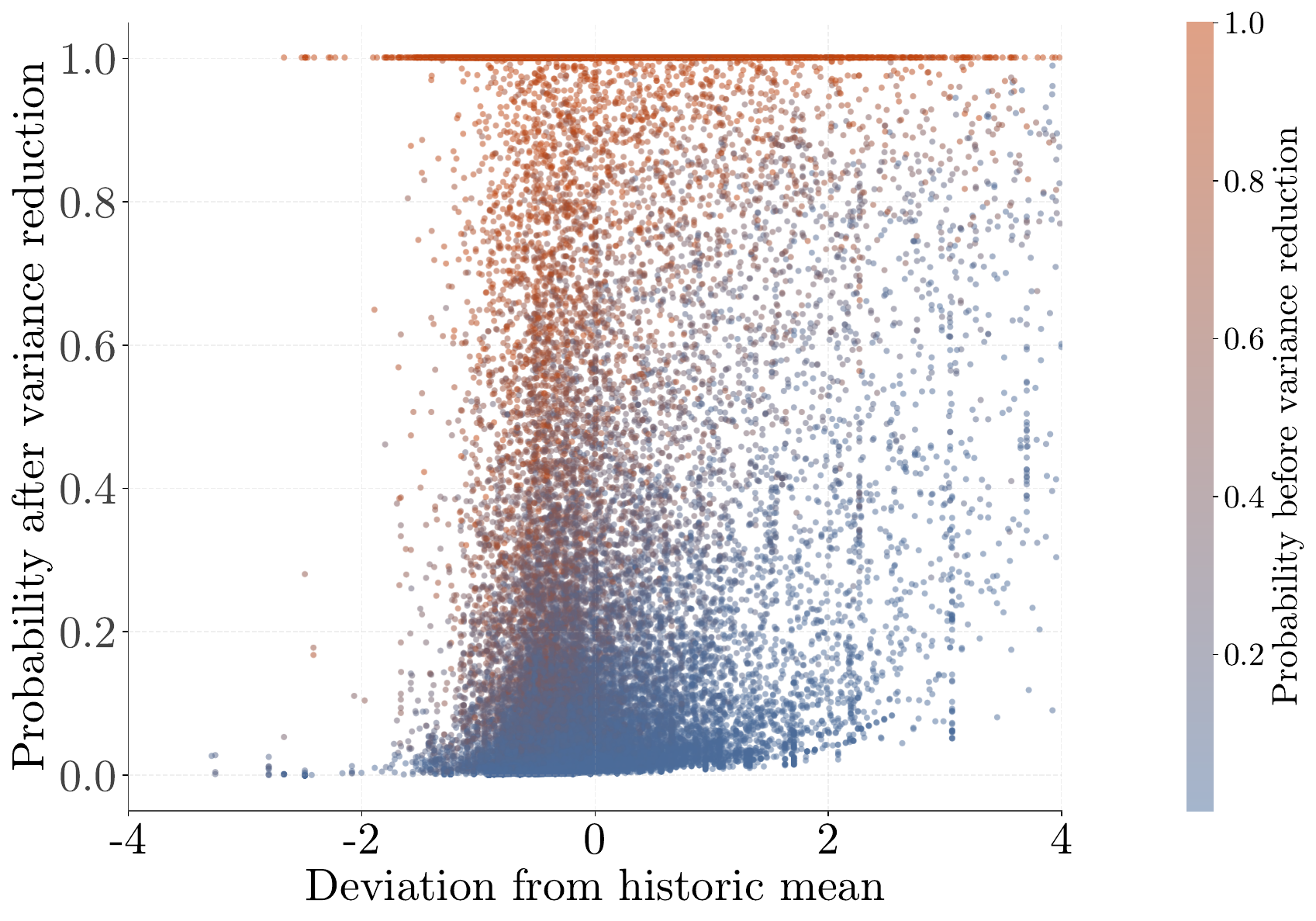}
    \caption{Variance-reduced probability of persistence-path write cycles (y-axis) relative to normalized event magnitude $w(t, e_n)$ (x-axis). Blue indicates events unlikely to trigger writes prior to variance reduction; orange indicates likely events. Variance-aware filtering increases retention probability only for statistically influential events.}
    \label{fig:skew-probability}
\end{figure}

To counter the error spikes caused by large-magnitude events in Equation~\ref{eq:variance-estimators}), we design inclusion rules that adjust \(p_n\) in inverse proportion to squared contributions \(w^2(t,e_n)\), i.e.
\begin{align}
    p_n = \sigma\!\left(
  \sigma^{-1}\!\left(\frac{\Lambda}{\hat{\lambda}(t_n)}\right) + \alpha\,\frac{w(t,e_n)-\mu_w}{\sigma_w}
\right),
\label{eq:variance-reduction}
\end{align}
where \(\sigma(x) = 1/(1+e^{-x})\) is the logistic function~\cite{berkson1944application}, and \(\mu_w\), \(\sigma_w\) denote historical means and standard deviations of contributions. Here, \(\alpha \in \mathbb{R}^+\) controls the trade-off between write-budget adherence and variance reduction. This mirrors classical importance sampling~\cite{tokdar2010importance, hammersley1964general, geweke1989bayesian} and is particularly effective under heavy-tailed event magnitudes~\cite{Vogel03042025}. Figures~\ref{fig:inclusion-probability-dist}--\ref{fig:skew-probability} demonstrate that this rule reallocates update probability toward informative events while stabilizing total write volume, and we refer the reader to Appendix~\ref{app:sensitivity-analysis} for a sensitivity analysis.

\begin{figure*}[t]
    \centering
    \includegraphics[width=\textwidth]{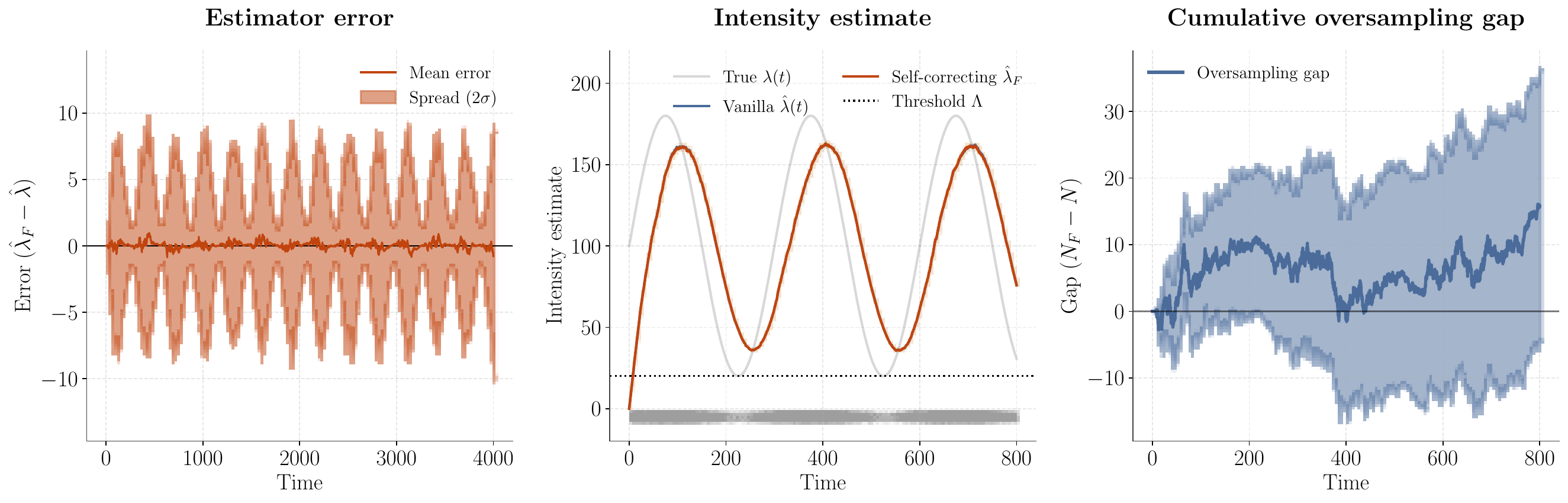}
    \caption{Left: Error bounds for $\hat{\lambda}_{\mathrm{F}}(t) - \hat{\lambda}(t)$ over $1000$ runs, illustrating cyclic self-correcting behavior. Center: True intensity of an arrival process with sampled events and corresponding estimators (blue line is hidden behind the red line). Right: Expected difference in write cycles under persistence-path control ($\mathcal{N}_{\mathrm{F}}$) and a vanilla in-memory control, with 95\% confidence intervals.}
    \label{fig:oversampling-updates}
\end{figure*}

\section{Filtering under Compute-Side Constraints}
\label{sec:compute-constrains}

A central challenge in operationalizing event filtering is the reliance on a continuously updated control state. Classical thinning assumes access to accurate arrival-rate estimates $\hat{\lambda}(t)$ or moments $(\mu_w, \sigma_w)$, maintained eagerly in compute-adjacent storage. However, maintaining high-frequency, coordination-sensitive control state reintroduces the very write pressure that thinning is meant to eliminate. If filtering decisions require per-event state updates, the control plane becomes a bottleneck even when data-plane updates are suppressed, nullifying the intended efficiency gains.

\paragraph{Design goal.} 
We require a filtering mechanism where decisions depend exclusively on persistence-backed state already maintained for feature computation, without additional in-memory state or coordination. Additionally, statistical correctness and predictable bounds on write rates must be preserved.

\subsection{Arrival-Rate Estimation as a Control Dependency}

To adjust $p_n$ dynamically, adaptive filtering requires an online estimate of the arrival intensity $\lambda(t)$ per entity. In practice, this is derived using Kalman filters~\cite{welch1995introduction} or Kernel Density Estimation (KDE) with exponential smoothing~\cite{rosenblat1956remarks,parzen1962estimation}:
\begin{equation}
    \hat{\lambda}(t) = \frac{1}{h}\sum_{e_n\in E(t)} \exp\left(-\frac{|t-t_n|}{h}\right),
    \label{eq:KDE-lambda}
\end{equation}
where $h>0$ is a \textit{bandwidth} controlling temporal responsiveness. In a streaming context, this admits a simple recurrence across timestamps $(t_n)_{n=1,2,\dots}$:
\[
  \hat{\lambda}(t_n) = \frac{\hat{\nu}(t_n)}{h}, \quad
  \hat{\nu}(t_n) = 1 + \exp\!\left(-\frac{t_n - t_{n-1}}{h}\right)\hat{\nu}(t_{n-1}).
\]

This estimator offers a low-variance, biased approximation of the true arrival intensity (see Appendix~\ref{sec:app-kde}) and is widely used in practice~\cite{lampe2011interactive, achar2019bus, bianchi2003kalman}. However, maintaining this estimate eagerly creates a circular dependency: the system must update compute-adjacent state just to decide if a disk-backed persistence update is necessary. Under high-throughput workloads, this feedback loop can dominate CPU time, serialization cost, and cache pressure.

\subsection{Persistence-Path Control via Filtered Estimators}

To break this dependency, we replace a full-stream KDE estimator with a \emph{filtered} counterpart:
\[
  \hat{\lambda}_\mathrm{F}(t_n) = \frac{1 
    + \exp\!\left(-(t_n - t_{n-1})/h\right)\hat{\nu}_{\mathrm{F}}(t_{n-1})}{h},
\]
where $\hat{\nu}_{\mathrm{F}}$ is updated exclusively on the persistence path, i.e.
\[
  \hat{\nu}_{\mathrm{F}}(t_{n})
  = \frac{Z_n}{p_n}
    + \exp\!\left(-\frac{t_n - t_{n-1}}{h}\right)\hat{\nu}_{\mathrm{F}}(t_{n-1}).
\]
only when a write operation is executed across persisted events $\mathcal{I} = \{e_n \in E : Z_n = 1\}$. Consequently, thinning probabilities $p_n$ in Equation~\eqref{eq:downsampling-prob} are derived using only disk-backed aggregates already required for feature maintenance, fully decoupling the process from in-memory control panes and synchronization. The trade-off is increased stochasticity, since $\hat{\lambda}_{\mathrm{F}}(t_n)$ is itself influenced by prior filtering decisions $Z_1, \dots, Z_{n-1}$. Crucially, this interdependence remains stable and analytically tractable.

\begin{remark}[Stability of filtered estimation]
\label{rem:martingale}
Define the normalized deviation
\[
    M_n = \frac{\hat{\lambda}_{\mathrm{F}}(t_n) - \hat{\lambda}(t_n)}{\exp\!\left(-t_n/h \right)}
\]
Then $\{M_n\}_{n \ge 0}$ is a martingale with respect to the natural filtration induced by event arrivals.
\end{remark}

The filtered estimator remains anchored in expectation to the full-stream KDE $\hat{\lambda}(t)$ in Equation~\eqref{eq:KDE-lambda}. This property motivates the term \emph{self-correcting}: if it overshoots, inclusion probabilities decrease, suppressing subsequent updates; when it undershoots, probabilities increase, restoring update frequency. Estimation errors cannot compound indefinitely, and formal proofs are provided in Appendix~\ref{app:double-stochastic-follows-correct-counts}.

Moreover, the variance introduced by filtered estimation provides an implicit safety guarantee through structural \textit{oversampling}.

\begin{remark}[Safety via oversampling]
\label{rem:oversampling}
Let $\mathcal{N}_{\mathrm{F}}$ and $\mathcal{N}$ denote the number of persistent updates triggered under filtered and full-stream control, using intensity estimates $\hat{\lambda}_{\mathrm{F}}(t)$ and $\hat{\lambda}(t)$ respectively. Then
\[
    \mathbb{E}[\mathcal{N}_{\mathrm{F}}] \;\ge\; \mathbb{E}[\mathcal{N}].
\]
\end{remark}

Persistence-path control may incur marginally higher write volume than an idealized in-memory design, but never fewer in expectation. As shown in Figure~\ref{fig:oversampling-updates}, oversampling is bounded, preserves statistical correctness, and can be managed by adjusting the global write budget $\Lambda$. 

\begin{figure*}[t]
    \centering
    \includegraphics[width=0.98\textwidth]{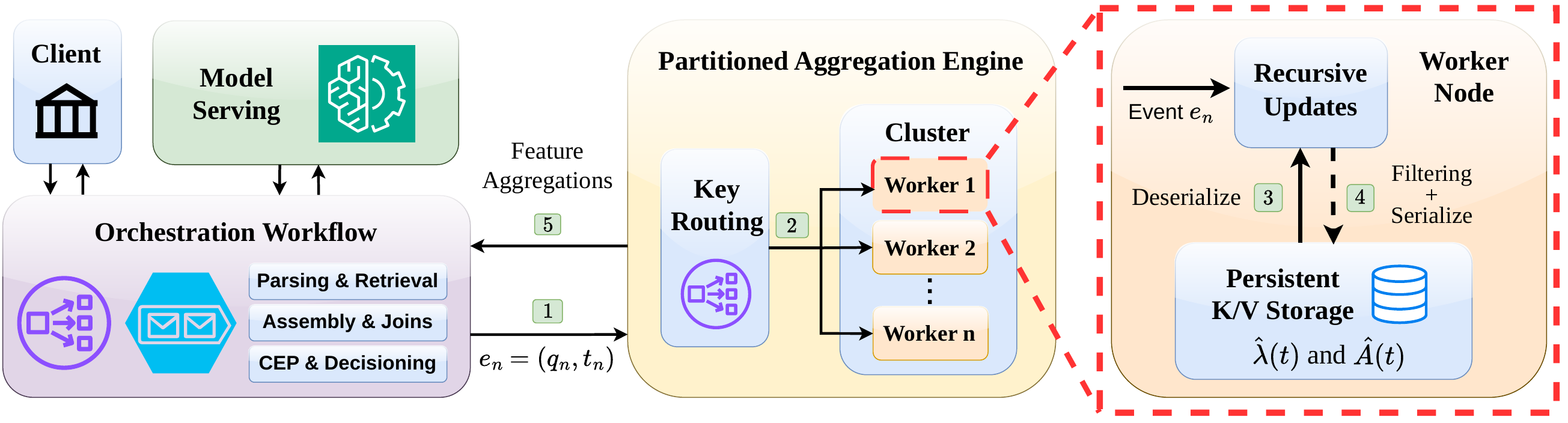}
    \caption{Logical architecture of a standard ML pipeline with persistence-path control embedded in the feature aggregation engine. A stream processing layer orchestrates feature retrieval and model inference, while state is maintained in partitioned key--value stores. (1) The orchestrator queries the engine for entity state; (2) requests are routed to the responsible partition, ensuring per-key ordering; (3) a worker retrieves feature aggregates and control statistics; (4) updates are probabilistically materialized via persistence-path control; (5) resulting features are returned to the orchestrator for downstream scoring.}
    \label{fig:system-architecture}
\end{figure*}

Taken together, results demonstrate that adaptive event filtering can be implemented without high-frequency in-memory control state. By relying exclusively on persistence-backed estimates, the system preserves unbiasedness, ensures predictable write-rate bounds, and control-plane computation scales with the write budget rather than the raw event rate. This aligns with the execution model of modern streaming engines, where only persisted state is checkpointed, replicated, and migrated during failure recovery or rebalancing. Appendix~\ref{app:double-stochastic-oversampling} provides formal proofs and additional analysis of oversampling dynamics.

\section{System Implementation}
\label{sec:system-implementation}

We describe a concrete system instantiating our probabilistic state management design. To isolate \textit{persistence-path control} as a reusable primitive, we focus on a standard execution unit: a worker operating over partitioned key-–value state. This allows us to reason about the marginal cost of persistence operations while maintaining compatibility with distributed execution.

\subsection{Component-Level Architecture}

Figure~\ref{fig:system-architecture} illustrates our design operating as an embedded feature aggregation component in a standard production-grade risk-scoring pipeline. The architecture consists of two primary layers:
\begin{itemize}
    \item A \textbf{Stream Processing Layer} orchestrates ingestion, feature retrieval, and model inference. This may include enrichment or \textit{Complex Event Processing} (CEP) pattern detection~\cite{zhao2020load, slo2020state}, and operates under strict latency constraints.
    \item A \textbf{Feature Aggregation Engine} maintains per-entity state in a distributed store, enabling recursive updates via partitioned key--value stores. Here, our mechanism intercepts the RMW cycle associated with feature updates.
\end{itemize}
Model inference is executed by a decoupled serving layer, which performs request routing, model selection, and stateless evaluation over the assembled feature vector, as is standard in modern online inference systems. In our partitioned system, for each event, a feature aggregation worker performs the following steps:
\begin{enumerate}
    \item Retrieves feature state and control statistics from storage.
    \item Materializes features for inference.
    \item Derives an inclusion probability over disk-backed estimates.
    \item Samples a Bernoulli decision.
    \item Executes a write-back only if selected.
\end{enumerate}
Crucially, inference is performed for \textit{every} event, while persistence updates are selectively suppressed.

The mechanism operates entirely within the feature computation layer: its control logic is stateless, requires no auxiliary in-memory control or cross-worker synchronization, and avoids control-plane amplification under skewed workloads. The design aligns with execution models of Apache Flink, Kafka Streams, and Spark Structured Streaming, where workers process disjoint key partitions with local state.

\subsection{Summary of Operating Assumptions}

We summarize the assumptions underlying the design and applicability of persistence-path control for feature-store workloads:
\begin{itemize}
    \item \textbf{Partitioned Storage:} State is organized per-entity and processed independently without cross-key coordination.
    \item \textbf{Recursive Aggregations:} We target decomposable feature updates of the form
    \[
    A(t_n) = g(A(t_{n-1}), e_n),
    \]
    for recursive operations $g$ including counts, sums or moments. Sequence-sensitive or pattern-based state (e.g., CEP operators) are out of scope.
    \item \textbf{Logical RMW:} Processing follows a retrieval, update, and optional persistence pattern. Optimizations like batching or async I/O must preserve these semantics.
    \item \textbf{Skewed Distributions:} We assume skewed workloads where a few keys drive most updates. Common condition essential for significant persistence reduction.
    \item \textbf{Approximation Tolerance:} Features tolerate controlled stochastic approximation, typical in domains like fraud detection where full inference coverage is prioritized.
\end{itemize}

\subsection{Controlled Execution Model}

We implement an execution model that instantiates the per-partition abstraction described above, under varying levels of concurrency, load, and key distribution. Our objective is to isolate the impact of probabilistic filtering on persistence costs, under storage and compute resource contention:
\begin{itemize}
    \item Events are replayed from Parquet-backed datasets using a containerized load generator based on Locust~\cite{Locust2024}.
    \item Requests are issued continuously and routed across worker instances via deterministic key partitioning, ensuring per-key ordering while enabling parallel processing.
\end{itemize}
Workers are deployed as independent, horizontally scaled containers on shared hardware, enforcing strict request–response execution with no internal batching. Thus, the system is subject to CPU contention, I/O interference, and skew-induced load imbalance across partitions. Each process orchestrates feature retrieval, probabilistic filtering, and optional persistence via a single logical RMW operations per event. This ensures that write frequency, IOPS, and latency remain directly observable. State management is handled through embedded RocksDB instances~\cite{dong2021rocksdb} with \textit{leveled} compaction and write-ahead logging, where updates are issued synchronously to SSD-backed storage to guarantee atomic, strongly ordered, and sequential execution per key.

\paragraph{Interpretation}
This execution model captures the fundamental unit of stateful processing in modern streaming systems: independent, partitioned workers operating over local key--value state. We thus preserve the persistence interface where RMW costs arise, while abstracting away runtime concerns such as operator scheduling or networked dataflows.

Because partitions execute independently, the measured effects of reduced write frequency directly translate to improved throughput, reduced contention, and delayed saturation under scale-out. These effects compose across partitions and integrate with common dataflow, check-pointing, and fault-tolerance mechanisms without requiring changes to upstream or downstream operators.

\section{Experimental Evaluation}
\label{sec:evaluation}

Our evaluation reflects streaming ML settings where all events are processed for inference, while only a subset of events is required to update persistent feature state. We separate concerns: system-level results isolate persistence overheads under varying execution conditions, while ML evaluation assesses the predictive impact of self-correcting persistence-path thinning. All results are reported with approximate 95\% confidence intervals, and experiments are reproducible and public.\footnote{\url{https://anonymous.4open.science/r/Decoupling-Inference-from-State-Updates-via-Probabilistic-Thinning-in-Low-Latency-Feature-Engines-0319}}

\subsection{Experimental Setup}

We report results on the datasets outlined in Table~\ref{tab:datasets} to capture distinct operational regimes:
\begin{itemize}
    \item \textbf{Financial Transaction Fraud}: A proprietary real-world transaction monitoring dataset, with strong key skew, continuous anomalous activity, and heavy-tailed transaction amount distributions.\footnote{Sourced for research purposes by Feedzai: \url{https://www.feedzai.com}}
    \item \textbf{IBM Synthetic Fraud Dataset}: Public benchmark exhibiting financial patterns with strong key skew and moderate heavy-tailed transaction values~\cite{altman2019synthesizingcreditcardtransactions}.
    \item \textbf{Edge-IIoTset}: Network intrusion data with bursty \textit{denial of service} patterns, temporal locality of anomalies, and relatively symmetric packet size distributions~\cite{ferrag2022edge}.
    \item \textbf{Wikipedia Vandalism}: User edit streams with weak key skew and balanced edit distributions, where anomalous behavior is associated with short-lived throwaway accounts, limiting the utility of long-lived temporal profiles~\cite{potthast2010crowdsourcing}.
\end{itemize}

Thus, we evaluate persistence-path control under target conditions (high key skew), assumption violations (e.g., weak skew or limited temporal structure), and across varying label magnitude imbalance. In all datasets, events are partitioned by a primary identifier, e.g. \emph{merchant}, \textit{user} or \textit{IP address}, and each key maintains temporal aggregations of activity patterns. Feature engineering is kept simple and production-representative. Persisted state consists exclusively of time-decayed aggregations, including counts, sums, and means, implemented via recursive decays as described in Section~\ref{sec:adaptive-event-filtering}. We use decay factors approximating windows of 1 minute, 1 hour, and 1, 30, 60, and 120 days. This captures heterogeneous temporal dynamics, from short-term bursts to long-term trends, while remaining compatible with constant-space KV-store updates.

Experiments run on dedicated AWS instances, with ingestion workers on \texttt{c6i.xlarge} nodes and load generation on \texttt{c5a.4xlarge} nodes. Components are co-located in the same region and \textit{Virtual Private Cloud} (VPC) to minimize network variability, and storage is backed by SSD EBS volumes.

\begin{table}[t]
    \centering
    \small
    \caption{Dataset characteristics. ``80\% Vol.'' denotes the percentage of keys responsible for 80\% of events; kurtosis refers to the aggregand distribution (e.g., transaction amounts).}
    \label{tab:datasets}
    \renewcommand{\arraystretch}{1.15}
    \setlength{\tabcolsep}{5.2pt}
    \begin{tabular}{lccccc}
    \toprule
    \textbf{Dataset} & \textbf{Events} & \textbf{Keys} & \textbf{Anomaly \%} & \textbf{80\% Vol.} & \textbf{Kurtosis}\\
    \midrule
    Fraud       & 11M   & 7K   & 0.05 & 4.1\% & 8 \\
    IBM         & 9M   & 7K   & 0.13 & 1.5\% & 3\\
    IIoTset     & 5M   & 800K & 40.01 & 0.7\% & 2\\
    Wikipedia   & 6K   & 3K  & 8.35 & 23.6\% & 2\\
    \bottomrule
    \end{tabular}
\end{table}

\begin{table*}[t]
    \centering
    \small
    \caption{Intrinsic efficiency metrics for different filtering strategies and baselines, on open-source IBM data. We use user-defined reference bounds for event volume processed, per minute and key. Full-stream filtering and periodic batching serve as idealized baselines that contextualize upper limits of achievable efficiency.}
    \label{tab:system-performance}
    \renewcommand{\arraystretch}{1.15}
    \setlength{\tabcolsep}{8.1pt}
    \ADLnullwide
    \begin{tabular}{lccccccccc}
        \toprule
        \textbf{Strategy} &
        \textbf{$\boldsymbol{\Lambda}$}&
        \textbf{Write}&
        \textbf{Throughput} &
        \multicolumn{3}{c}{\textbf{Latency} (ms)} &
        \multicolumn{3}{c}{\textbf{Fixed Throughput} @ 200 Ev/s}\\ 
        \cmidrule(lr){5-7} \cmidrule(lr){8-10}
        & (Ev/m) & (\%) & (TPS) & \textbf{Avg} & \textbf{p95} & \textbf{p99.99} & \textbf{WAF} & \textbf{Bps} & \textbf{Util (\%)} \\
        \midrule

        \textbf{Unfiltered} 
        & -- & 100.00 & 226.82 $\pm$ 0.21 & 4.36 $\pm$ 0.00 & 5.00 & 10.00 & 2.6 & 3910.54 $\pm$ 17.45 & 48 \\
        \midrule

        \textbf{Persistence-Path} 
        & $0.001$ & 5.91 $\pm$ 0.05 & 621.05 $\pm$ 3.15 & 1.56 $\pm$ 0.01 & 4.00 &  6.50 & 1.7 & 247.07 $\pm$ 3.24 & 3 \\
        & $0.005$ & 25.65 $\pm$ 0.07 & 454.90 $\pm$ 1.80 & 2.16 $\pm$ 0.00 & 4.00 &  8.25 & 2.3 & 1029.48 $\pm$ 3.81 & 13 \\
        & $0.010$ & 44.70 $\pm$ 0.04 & 352.57 $\pm$ 0.41 & 2.79 $\pm$ 0.00 & 4.33 & 8.80 & 2.3 & 1782.59 $\pm$ 5.87 & 22 \\
        & $0.050$ & 83.94 $\pm$ 0.02 & 252.00 $\pm$ 0.14 & 3.91 $\pm$ 0.00 & 5.00 & 9.80 & 2.7 & 3317.67 $\pm$ 6.21 & 41 \\
        & $0.100$ & 91.70 $\pm$ 0.02 & 239.41 $\pm$ 0.15 & 4.13 $\pm$ 0.00& 5.00 & 9.93 & 2.6 & 3625.38 $\pm$ 8.68 & 45 \\
        & $1.000$ & 100.00 $\pm$ 0.00 & 226.59 $\pm$ 0.27 & 4.37 $\pm$ 0.00 & 5.00 & 10.00 & 2.7 & 3947.12 $\pm$ 7.30 & 49 \\
        \midrule

        \textbf{Full-Stream} 
        & $0.010$ & 13.57 $\pm$ 0.78& 510.78 $\pm$ 6.19 & 1.91 $\pm$ 0.02 & 4.00 & 6.40 & 2.1 & 781.32 $\pm$ 7.25 & 8 \\
        & $0.050$ & 55.24 $\pm$ 0.04& 318.74 $\pm$ 0.35 & 3.09 $\pm$ 0.00 & 4.73 & 9.13 & 2.3 & 3269.48 $\pm$ 2.90 & 40 \\
        & $0.100$ & 73.34 $\pm$ 0.04& 274.31 $\pm$ 0.21 & 3.60 $\pm$ 0.00 & 5.00 & 9.33 & 2.5 & 3755.13 $\pm$ 3.14 & 46 \\
        & $1.000$ & 99.84 $\pm$ 0.03 & 227.03 $\pm$ 0.43 & 4.36 $\pm$ 0.00 & 5.00 & 9.93 & 2.6 & 3970.73 $\pm$ 184.31 & 49 \\
        \midrule

        \textbf{Fixed-Rate}
            & -- & 14.98 $\pm$ 0.08 & 536.68 $\pm$ 2.04 & 1.82 $\pm$ $0.01$ & 4.00 & 7.00 & 2.0 & 559.85 $\pm$ 3.67 & 7 \\
            & -- & 44.94 $\pm$ 0.16 & 360.56 $\pm$ 2.77 & 2.73 $\pm$ 0.02 & 4.00 & 9.25 & 2.0 & 1773.27 $\pm$ 6.79 & 22 \\
        \midrule

        \textbf{Periodic Batching}
            & -- & -- & 703.12 $\pm$ 0.47 & 1.38 $\pm$ 0.00 & 1.00 & 6.00 & 2.4 & 128.84 $\pm$ 2.41 & 1 \\
        \bottomrule
    \end{tabular}
\end{table*}

\subsection{Filtering Strategies and Baselines}

We compare our design against a range of strategies that isolate different system trade-offs, using arrival-intensity estimates $\hat{\lambda}(t)$ for filtering decisions computed with the KDE estimator in Equation~\eqref{eq:KDE-lambda}. All strategies enforce a user-defined upper bound $\Lambda$ on the expected number of persistent write operations per minute and key. We evaluate the following strategies:
\begin{enumerate}
    \item \textbf{Persistence-Path Control.}
    Filtering decisions depend solely on control statistics updated along the persistence path, without auxiliary in-memory control state.
    \item \textbf{Persistence-Path Control + Variance Reduction.}
    Inclusion probabilities incorporate the variance-aware formulation in Equation~\eqref{eq:variance-reduction}, evaluated specifically for downstream ML fidelity and feature approximation quality.
\end{enumerate}
As benchmarks, we consider the following reference baselines and commonly used techniques for controlling write pressure:
\begin{enumerate}
    \item \textbf{Full-Stream Control.}
    Reference baseline with compute-adjacent control state, enabling filtering decisions without serialization overheads.
    \item \textbf{Naive Fixed-Rate Filtering.}
    Stateless random persistent updates using a fixed global probability, independent of key activity or arrival rate.
    \item \textbf{Periodic Batching.}
    Per-key buffering with periodic flushes, introducing feature staleness and bursty I/O. We use a small buffer ($100$ events) to approximate an upper bound on batching efficiency under minimal delay.
\end{enumerate}

\paragraph{Excluded Baselines.}
We exclude reservoir sampling~\cite{reservoir_sampling} and its weighted variants~\cite{efraimidis2006weighted}. These techniques maintain a fixed-size sample per key, requiring explicit in-memory buffers whose size must be chosen a priori. Under skewed key distributions, this leads to unbounded memory growth or biased eviction behavior, making reservoir sampling unsuitable for high-cardinality feature maintenance with strict write-rate constraints.

\subsection{Intrinsic Execution}
\label{sec:systems-eval}

We first evaluate system-level efficiency in a controlled single-partition setting that isolates the direct relationship between persistence frequency and worker-level performance. This captures intrinsic properties of the RMW execution model underlying modern streaming systems, abstracting away cross-partition contention, coordination, or scheduling effects. We use the IBM dataset. Since system behavior is primarily driven by event rate and persistence frequency rather than feature semantics, results are qualitatively consistent across datasets. We use two configurations:
\begin{itemize}
    \item Closed-loop load generation~\cite{schroeder2006open} to measure \textbf{client-side} peak throughput and latency, issuing each subsequent event only after receiving a response, and 
    \item Fixed-rate input ($200$ events/sec) to isolate \textbf{system-side} resource utilization.
\end{itemize}
We report on:
(i) write frequency,
(ii) throughput,
(iii) end-to-end latency,
(iv) write amplification,
(v) sustained storage bandwidth, and
(vi) disk utilization.
Figure~\ref{fig:teaserMain} illustrates the throughput--volume trade-off, and Table~\ref{tab:system-performance} summarizes intrinsic performance results.

Across all system-level metrics, \emph{persistence-path control} achieves substantial reductions in write amplification, latency, and disk utilization compared to reference baselines, while increasing throughput, despite avoiding in-memory control state. Also, as $\Lambda \rightarrow 1$, performance metrics converge to the unfiltered baseline, confirming negligible overhead.

\paragraph{Client-Side Metrics.}
Unfiltered processing yields the minimum throughput (226.8 TPS), while decreasing the write budget $\Lambda$ results in near-linear scaling: at write budget $\Lambda=0.001$, throughput increases $2.7\times$ and average latency drops $64\%$. Stable $p99.99$ tail latencies indicate that aggressive filtering mitigates I/O-induced head-of-line blocking in synchronous RMW execution, ultimately improving ML utility (Section~\ref{sec:ml-eval}).

\paragraph{System-Side Metrics.}
Filtering reduces WAF from 2.6 to 1.7. This aligns with LSM-tree dynamics, where lower ingestion rates suppress compaction frequency and background amplification~\cite{dong2021rocksdb}, extending SSD endurance. The non-linear rise in disk utilization at higher $\Lambda$ reflects LSM-tree compaction thresholds, where background maintenance amplifies I/O disproportionately~\cite{dayan2017monkey,lu2017wisckey}; filtering effectively keeps the system below these saturation points.

Noticeably, periodic batching achieves high throughput by amortizing I/O, but at the cost of feature staleness and streaming ML inference degradation~\cite{DaiZDZX19} (see results in Section~\ref{sec:ml-eval}). The strong tail-latency results observed reflect small buffers and single-threaded execution; larger batch sizes or multi-partition deployments would induce bursty I/O and unbounded staleness.

\begin{figure*}[t]
    \centering
    \includegraphics[width=\textwidth]{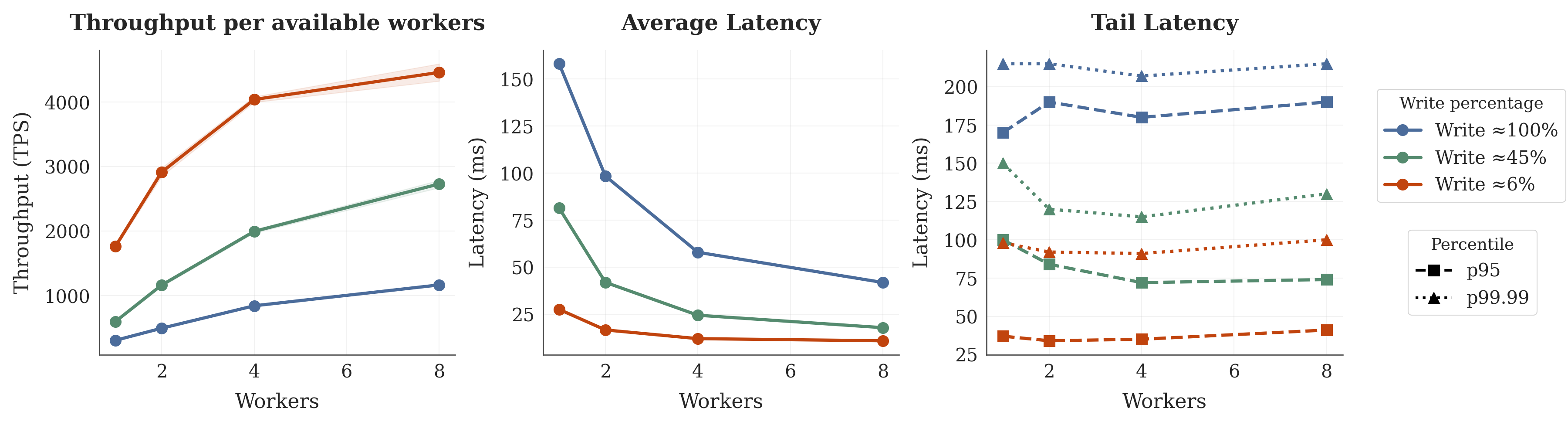}
    \caption{Concurrent execution and contention evaluation under horizontally scaled partitioned workers. Events are routed to worker instances via deterministic key partitioning, preserving per-key ordering while enabling parallel execution over isolated RocksDB-backed state. Left: Throughput under increasing worker parallelism. Center: average latency under contention. Right: tail-latency behavior. Relative gains from persistence-path control remain stable under increasing CPU and I/O contention.}
    \label{fig:concurrent}
\end{figure*}

\subsection{Operational Scalability}

We next demonstrate that the relative benefits of \emph{persistence-path control} remain stable under operational conditions common in production streaming systems, including concurrency, asynchronous execution, and workload variability. In particular, we study: (i) concurrent execution under increasing parallelism, (ii) sensitivity to skewed key distributions, (iii) long-running execution stability, and (iv) behavior near saturation points.

\paragraph{Concurrent Execution and Contention.}
Figure~\ref{fig:concurrent} evaluates throughput and latency while horizontally scaling independent partition workers across containerized RocksDB-backed instances. Events are injected concurrently across workers using deterministic key-to-worker routing, following the partitioned aggregation architecture illustrated in Figure~\ref{fig:system-architecture}, while increasing aggregate load to expose the system to significant CPU and I/O contention.

We observe throughput scaling consistently with worker parallelism, while average latency decreases despite higher aggregate load. Importantly, the relative gains from filtering remain stable under increasing contention. Stable tail latencies further suggest that reducing persistence frequency mitigates contention-induced queueing and head-of-line blocking during concurrent execution.

\paragraph{Sensitivity to Skew, Stability, and Saturation.}
Table~\ref{tab:operational-scalability} summarizes worker-level analyses under this partitioned execution model, across varying workload regimes. To study skew sensitivity, we reduce key-frequency imbalance by removing transactions associated with high-volume entities, while preserving overall workload structure. This materially changes the realized write percentage under identical filtering budgets. Nevertheless, throughput and latency remain anchored to persistence frequency rather than the key distribution, suggesting that the dominant execution cost remains the number of completed RMW cycles.

We further evaluate long-running execution by extending workload duration from $5$ to $50$ minutes across filtering regimes, under stabler request volumes. Here, throughput and latency remain effectively unchanged over time, indicating stable behavior and absence of progressive degradation under sustained operation.

Finally, we evaluate saturation behavior by progressively increasing the intensity of asynchronous request streams until latency collapse (>$500$ms), in increments of $50$ events/sec to estimate failure regions. The sustainable throughput prior to back-pressure onset increases substantially as filtering becomes more aggressive, with the failure threshold rising from approximately $200$ events/sec to over $1350$ events/sec at the lowest write rates. This indicates that reducing persistence pressure delays storage saturation and allows substantially higher inference throughput before contention dominates execution. More broadly, persistence-path control expands the feasible operating region of streaming inference systems by enabling higher event throughput before storage saturation or back-pressure compromise latency.

\begin{table}[h!]
\centering
\caption{System evaluation under key skew variation, long-running execution, and saturation analysis.}
\label{tab:operational-scalability}

\renewcommand{\arraystretch}{1.07}
\setlength{\tabcolsep}{8pt}
\small

\begin{tabular}{lcccc}
    \toprule
    \textbf{Experiment} &
    \textbf{Throughput} &
    \multicolumn{3}{c}{\textbf{Latency} (ms)} \\ 
    \cmidrule(lr){3-5}
    & (TPS) & \textbf{Avg} & \textbf{p95} & \textbf{p99.99} \\
    \midrule
    
    \multicolumn{4}{l}{\textbf{Sensitivity to Key Skew}} \\
    \midrule

    \multicolumn{4}{l}{\textbf{10\% keys $\rightarrow$ 80\% volume}} \\
    \hspace{2mm}Write $\approx$70\% & 270.54 $\pm$ 1.35 & 3.65 & 5 & 10 \\
    \hspace{2mm}Write $\approx$85\% & 241.27 $\pm$ 1.07 & 4.10 & 5 & 10 \\
    \hspace{2mm}Write $\approx$97\% & 220.76 $\pm$ 1.69 & 4.48 & 5 & 10 \\

    \multicolumn{4}{l}{\textbf{5\% keys $\rightarrow$ 80\% volume}} \\
    \hspace{2mm}Write $\approx$41\% & 354.38 $\pm$ 1.13 & 2.77 & 5 & 8.5 \\
    \hspace{2mm}Write $\approx$61\% & 294.57 $\pm$ 1.37 & 3.35 & 5 & 9.25 \\
    \hspace{2mm}Write $\approx$87\% & 241.10 $\pm$ 0.52 & 4.10 & 5 & 10 \\

    \midrule

    \multicolumn{4}{l}{\textbf{Long-Running Stability}} \\
    \midrule

    \multicolumn{4}{l}{\textbf{Write $\approx$100\%}} \\
    \hspace{2mm}5 Minutes & 221.21 $\pm$ 0.54 & 4.47 & 5 & 9.75 \\
    \hspace{2mm}10 Minutes & 221.44 $\pm$ 0.87 & 4.47 & 5 & 9.75 \\
    \hspace{2mm}50 Minutes & 220.72 $\pm$ 0.63 & 4.48 & 5 & 10 \\

    \multicolumn{4}{l}{\textbf{Write $\approx$45\%}} \\
    \hspace{2mm}5 Minutes & 343.59 $\pm$ 1.42 & 2.86 & 5 & 8.5 \\
    \hspace{2mm}10 Minutes & 342.96 $\pm$ 1.73 & 2.87 & 5 & 9 \\
    \hspace{2mm}50 Minutes & 342.15 $\pm$ 1.06 & 2.88 & 5 & 9 \\

    \multicolumn{4}{l}{\textbf{Write $\approx$6\%}} \\
    \hspace{2mm}5 Minutes & 559.88 $\pm$ 1.77 & 1.74 & 4 & 5.5 \\
    \hspace{2mm}10 Minutes & 559.00 $\pm$ 2.52 & 1.74 & 4 & 6 \\
    \hspace{2mm}50 Minutes & 557.01 $\pm$ 1.17 & 1.75 & 4 & 6 \\

    \midrule

    \multicolumn{4}{l}{\textbf{Backpressure and Saturation}} \\
    \midrule

    \textbf{Write $\approx$100\%} & \textbf{Failure Thr.} & & & \\
    \hspace{2mm}100\% & 200-250 & -- & -- & -- \\

    \textbf{Filtered Execution} & & & & \\
    \hspace{2mm}Write $\approx$45\% & 350-400 & -- & -- & --\\
    \hspace{2mm}Write $\approx$26\% & 600-650 & -- & -- & --\\
    \hspace{2mm}Write $\approx$6\% & 1350-1400 & -- & -- & --\\

\bottomrule
\end{tabular}
\end{table}

\begin{table*}[h]
    \centering
    \small
    \caption{Downstream ML performance across datasets and experiments, under varying filtering strategies. As we relax human-defined RMW budgets, we measure write persistence operations and model recall at fixed FPR levels.}
    \label{tab:combined-performance}
    \renewcommand{\arraystretch}{1.15}
    \setlength{\tabcolsep}{8.45pt}
    \begin{tabular}{lcccccccc}
        \toprule
        \textbf{Strategy} & 
        \multicolumn{2}{c}{\textbf{Real Fraud}} &
        \multicolumn{2}{c}{\textbf{IBM Fraud}} &
        \multicolumn{2}{c}{\textbf{Edge-IIoT}} &
        \multicolumn{2}{c}{\textbf{Wikipedia}} \\
        \cmidrule(lr){2-3} \cmidrule(lr){4-5} \cmidrule(lr){6-7} \cmidrule(lr){8-9}
        & Write \% & Recall $\Delta$ & Write \% & Recall $\Delta$ & Write \% & Recall $\Delta$ & Write \% & Recall $\Delta$ \\
        \midrule
        \textbf{Unfiltered}
        & $100.00$ & $0.00$ & $100.00$ & $0.00$ & $100.00$ & $0.00$ & $100.00$ & $0.00$ \\
        \midrule

        \textbf{Persistence-Path}
        & $0.84$  & $-5.61 \pm 5.22$ & $33.34$ & $-2.94 \pm 0.27$ & $16.42$ & $-10.29 \pm 3.03$ & $87.11$ & $+0.05 \pm 0.35$\\
        & $3.60$  & $+7.06 \pm 2.53$ & $45.14$ & $-1.11 \pm 0.35$ & $27.51$ & $-7.19 \pm 5.79$  & $88.55$ & $+0.09 \pm 0.28$\\
        & $6.67$  & $+7.83 \pm 3.17$ & $54.03$ & $-0.40 \pm 0.39$ & $46.68$ & $+0.81 \pm 10.28$ & $89.88$ & $+0.05 \pm 0.28$\\
        & $28.11$ & $+4.55 \pm 5.03$ & $78.23$ & $-0.18 \pm 0.31$ & $49.51$ & $+2.46 \pm 11.78$ & $92.14$ & $-0.05 \pm 0.24$\\
        & $53.19$ & $+2.57 \pm 4.18$ & $90.26$ & $+0.45 \pm 0.26$ & $62.42$ & $+8.73 \pm 12.75$ & $94.19$ & $-0.17 \pm 0.26$\\
        \midrule

        \textbf{Persistence-Path + VR}
        & $1.10$  & $-5.21 \pm 5.94$ & $34.35$ & $-1.81 \pm 0.26$ & $16.89$ & $-10.91 \pm 0.02$ & $86.68$ & $-0.11 \pm 0.34$\\
        & $4.00$  & $+5.67 \pm 3.07$ & $46.30$ & $-0.98 \pm 0.34$ & $28.39$ & $-11.15 \pm 0.03$ & $88.33$ & $+0.02 \pm 0.27$\\
        & $7.23$  & $+6.00 \pm 3.76$ & $55.06$ & $+0.05 \pm 0.34$ & $47.67$ & $+0.78 \pm 11.31$ & $89.80$ & $+0.08 \pm 0.18$\\
        & $29.00$ & $+5.05 \pm 4.78$ & $78.62$ & $+0.09 \pm 0.41$ & $50.51$ & $+5.54 \pm 11.74$ & $92.07$ & $+0.11 \pm 0.30$\\
        & $53.52$ & $+0.45 \pm 2.77$ & $89.97$ & $+0.15 \pm 0.27$ & $62.73$ & $+6.74 \pm 12.40$ & $94.13$ & $+0.05 \pm 0.35$\\
        \midrule

        \textbf{Full-Stream}
        & $0.56$  & $-0.27 \pm 6.08$ & $32.21$ & $-2.77 \pm 0.29$ & $17.82$ & $-13.17 \pm 1.51$ & $85.22$ & $-0.21 \pm 0.40$\\
        & $2.80$  & $+4.07 \pm 5.91$ & $48.34$ & $-1.05 \pm 0.28$ & $32.14$ & $-8.33 \pm 5.97$ & $86.56$ & $+0.26 \pm 0.33$\\
        & $5.60$  & $+4.49 \pm 8.53$ & $58.60$ & $-0.10 \pm 0.29$ & $45.63$ & $+0.32 \pm 11.95$ & $87.72$ & $+0.24 \pm 0.35$\\
        & $27.90$ & $+3.98 \pm 4.39$ & $84.15$ & $+0.10 \pm 0.42$ & $51.04$ & $+2.79 \pm 11.30$ & $92.07$ & $+0.07 \pm 0.31$\\
        & $55.90$ & $+2.14 \pm 4.51$ & $95.52$ & $+0.12 \pm 0.34$ & $65.56$ & $+4.89 \pm 12.80$ & $94.14$ & $+0.22 \pm 0.38$\\
        \midrule

        \textbf{Fixed-Rate}
        & $1.00$ & $+1.63 \pm 3.74$ & $33.11$ & $-3.27 \pm 0.31$ & $16.32$ & $-11.05 \pm 4.79$ & $87.01$ & $-0.20 \pm 0.24$\\
        & $4.00$ & $+2.77 \pm 3.90$ & $54.36$ & $-1.76 \pm 0.18$ & $28.02$ & $-11.09 \pm 6.91$ & $92.31$ & $-0.40 \pm 0.21$\\
        & $7.00$ & $+2.53 \pm 4.64$ & $94.92$ & $-0.19 \pm 0.36$ & $45.87$ & $-11.17 \pm 3.09$ & $94.91$ & $-0.02 \pm 0.31$\\
        \midrule

        \textbf{Periodic Batching}
        & - & $-6.10 \pm 1.30$ & - & $-1.30 \pm 0.25$ & - & $-6.19 \pm 4.95$ & - & $-0.47 \pm 0.30$\\

        \bottomrule
    \end{tabular}
\end{table*}

\subsection{Downstream ML Utility}
\label{sec:ml-eval}

\begin{figure*}[t]
    \centering
    \includegraphics[width=\textwidth]{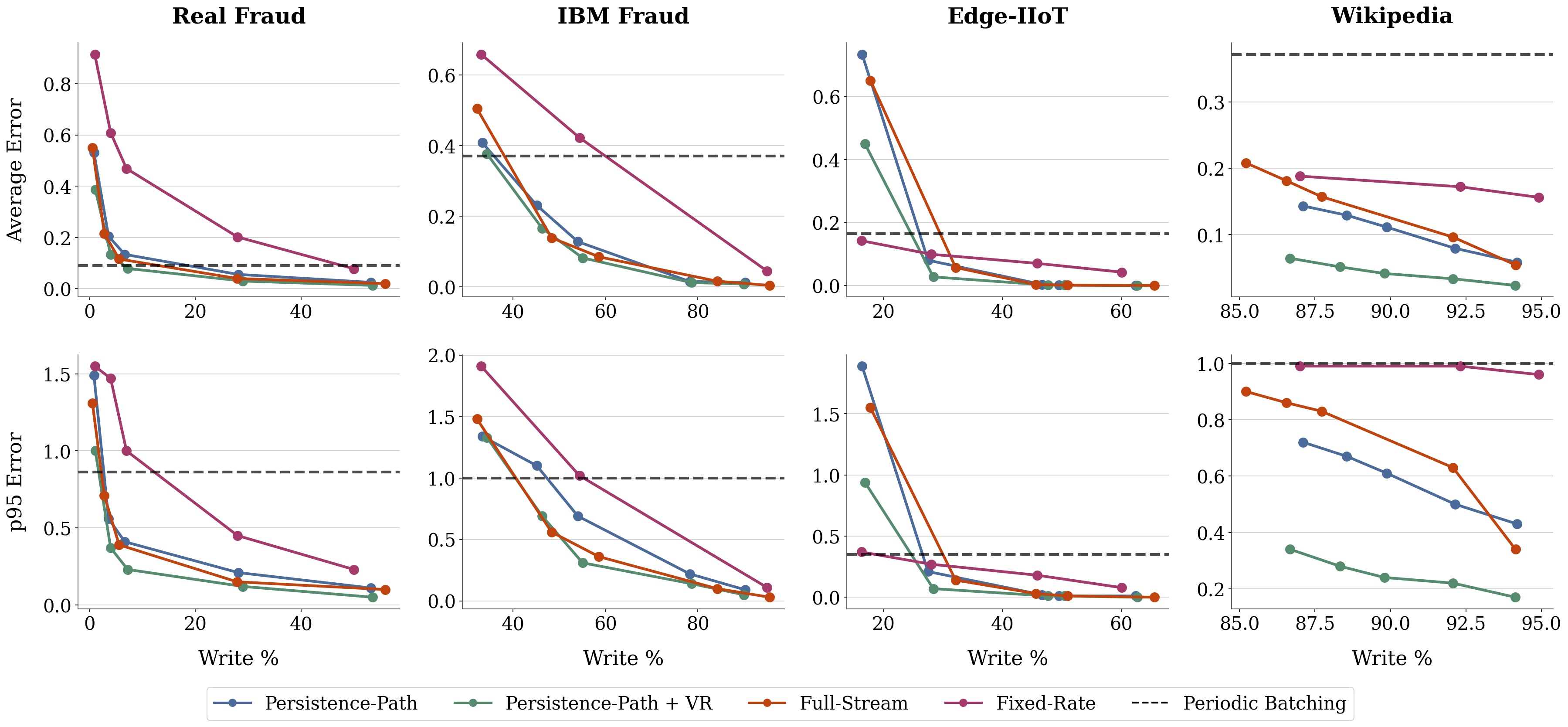}
    \caption{Sum feature aggregation approximation error (average and 95th percentile) across datasets, under varying filtering strategies and write persistence operations.}
    \label{fig:combined-error}
\end{figure*}

We next assess whether adaptive filtering:
(i) degrades downstream ML performance, and
(ii) preserves sufficient information to approximate full-stream feature aggregates.

Using the datasets and workload regimes in Table~\ref{tab:datasets}, we vary the write budget $\Lambda$ to aling \emph{write percentages} across strategies. These \textbf{differ by datasets due to event rate and volume}. We use multiple end-to-end simulations with temporal train--test splits, evaluating on \emph{all} test events, including those that did not trigger persistent updates. This reflects realistic deployment conditions where inference operates on potentially stale profiles. To isolate the effect of filtering, models use only features derived from aggregations in persistent state; stateless attributes such as timestamps and categorical encodings are excluded. Thus, we directly attribute changes in ML utility to the fidelity of maintained state. Finally, given the stochastic nature of filtering and ML training and evaluation, our analysis prioritizes consistent trends over individual point estimates.

\paragraph{Effect on ML Performance.}
In Table~\ref{tab:combined-performance} we report the write volume and recall \textit{differential} relative to an unfiltered baseline at a fixed \textit{False Positive Rate} (FPR) of $1\%$, a standard operating point in anomaly detection~\cite{perez2023locally}. Across datasets and operating regimes, \emph{persistence-path control largely preserves} the performance of unfiltered pipelines despite substantial reductions in persistence volume. At moderate filtering levels (e.g., $3$--$7\%$ write rates in transactional settings), recall improvements are sometimes observed, while aggressive filtering eventually degrades performance. This yields a \emph{non-monotonic trade-off}, with a broad operating region where substantial write reduction is achieved with limited or no loss in predictive quality. While recall improvements are observed in several workloads, other datasets primarily exhibit performance preservation under substantial write reduction, indicating that the benefits of adaptive filtering are workload-dependent rather than uniform.

We attribute performance improvements to implicit regularization induced by probabilistic thinning, i.e. filtering reduces the dominance of high-velocity keys, whiles stochastic suppression can inject controlled noise into feature construction. This is consistent with known effects in stochastic regularization~\cite{tikhonov,dropout,noiseinjection,krueger2017zoneoutregularizingrnnsrandomly}. However, we emphasize that improvements are workload-dependent rather than universal.

\paragraph{Approximation Fidelity and Variance Reduction.}
In Figure~\ref{fig:combined-error} we show the feature approximation error measured on \emph{sum aggregates}, which are particularly sensitive to missed updates and therefore represent a worst-case proxy for fidelity. Across datasets, approximation error decreases monotonically with increasing write volume. Even under substantial filtering, errors remain bounded, indicating that recursive aggregates degrade gracefully under stochastic update suppression. Furthermore, variance-aware filtering (green lines) consistently improves approximation fidelity relative to standard persistence-path control (blue lines), at comparable write rates. Notably, persistence-path control remains competitive with the full-stream in-memory benchmark (orange lines), and in several regimes achieves comparable or lower approximation error despite relying exclusively on persistence-backed control statistics.

\paragraph{Baselines.}
Full-stream control achieves similar recall and approximation quality to persistence-path control, suggesting that persistence-backed control statistics are sufficient to drive effective filtering decisions without requiring high-frequency in-memory state or coordination. Also, naive fixed-rate filtering performs inconsistently across datasets, since it disproportionately under-samples high-activity keys, leading to degraded feature fidelity and weaker performance recall. Finally, periodic batching introduces systematic degradation in ML performance and feature aggregation accuracy across datasets, delaying state updates and inducing feature staleness.

\section{Discussion}
\label{sec:discussion}

We have demonstrated that the tight coupling between inference and persistent state updates in streaming ML systems is not fundamental. By treating persistence updates as probabilistic events, rather than mandatory side effects of inference, we show that it is possible to substantially reduce write frequency, serialization overhead, and storage I/O while largely preserving (and in some workloads improving) downstream ML utility. Under skewed workloads, the marginal statistical value of successive updates for high-activity entities diminishes rapidly, whereas their systems cost grows linearly.

The \emph{persistence-path control} design avoids high-frequency in-memory control planes, coordination across workers, and additional fault-tolerance mechanisms. Despite operating on filtered control statistics, the resulting stochastic process is self-correcting, preserves unbiasedness, and enforces predictable write-rate bounds with only bounded oversampling. In our experiments, we show that persistence-path control remains competitive with full-stream in-memory coordination while operating exclusively on persistence-backed statistics. As a result, control-plane overhead scales with the write budget rather than the raw event rate, aligning naturally with the execution and recovery models of modern streaming engines.

From a broader systems perspective, our approach can be interpreted as introducing a probabilistic control layer within stateful stream processing. This is complementary to traditional complex event processing (CEP) and streaming dataflow systems, where operators typically assume deterministic state transitions. In contrast, persistence-path control relaxes this assumption by allowing controlled stochasticity in state updates, while preserving statistical correctness at the aggregate level. More generally, the results suggest that certain classes of stateful streaming workloads may tolerate substantially reduced persistence intensity before approximation effects materially impact downstream inference quality.

While the performance gains are significant, several architectural considerations and broader implications warrant discussion. First, the benefits of probabilistic filtering are workload-dependent: some datasets exhibit measurable improvements in downstream ML utility, while others primarily maintain baseline performance under substantial write reduction. Second, the stochastic nature of filtering introduces bounded approximation error in maintained state, which may or may not be acceptable depending on application requirements. In domains such as fraud detection, these errors are often secondary to system responsiveness under load; however, in contexts involving compliance, financial auditing, or legal reporting, discarding updates may not be permissible. In such scenarios, our architecture can be deployed in a \emph{tiered fashion}~\cite{10488241}: applying probabilistic filtering for low-latency inference while maintaining an unfiltered, strictly consistent path for auditability.

\paragraph{Future Extensions}
Promising directions for future research include:
\begin{itemize}
    \item Integrating streaming sketches and probabilistic data structures that are explicitly aware of probabilistic downsampling. This would enable richer feature profiles (e.g., quantiles, distinct counts) while maintaining statistical guarantees under filtered updates.
    \item Developing principled approaches to variance reduction, including analytical calibration of parameter $\alpha$ in Equation~\ref{eq:variance-reduction} under different data distributions and feature objectives, as well as alternative formulations that better prioritize high-impact events when low approximation error is required.
    \item Extending persistence-path control to more heterogeneous streaming deployments involving shared storage layers, multi-tenant contention, and distributed operator coordination.
\end{itemize}

\bibliographystyle{ACM-Reference-Format}
\bibliography{references}

\appendix

\section{Properties of Filtered Estimators}
\label{app:estimator-proofs}

Let $\mathcal{E}$ define a realization of fixed size from a Marked Inhomogeneous Counting Process
\[
    \mathcal{E} = \{ e_1, e_2, \dots, e_N \}, \quad e_n = (q_n, t_n),
\]
where $t_n \in \mathbb{R}^+$ are arrival times drawn from a time-varying intensity function $\lambda(t)$, with $t_N=t$. And consider the aggregate
\[
A(t)=\sum_{n=1,...,N} w(t, e_n).
\]
for an arbitrary function $w(t, e_n)$ at time $t>0$.

\subsection{The Estimator is Unbiased}
Let $p_n\in(0,1]$ for $n=1\dots, N$ be a sequence of random variables and $Z_n\sim \text{Bernoulli}(p_n)$. We show that:
\begin{equation*}
		\hat{A}(t)=\sum_{n=1,...,N} w(t, e_n) \cdot  \frac{Z_n}{p_n},
\end{equation*} 
is an unbiased estimator of $A(t)$, assuming that $w(t, e_n)$ is a fixed sequence over $1,\dots, N$ independent of $p_n$, for all $n\leq N$.

\begin{proof}
Let
\[
 \mathcal{F}_{n-1} = \sigma(Z_1,\dots,Z_{n-1})
\]
denote the natural filtration containing Bernoulli trials information available immediately before event $e_n$, such that $p_n$ is is predictable. Note that
\[
\mathbb{E}\left[\hat{A}(t)\right]
=
\mathbb{E}\left[\sum_{n=1,...,N} w(t, e_n) \cdot  \frac{Z_n}{p_n}\right]
=
\sum_{n=1,...,N} w(t, e_n) \cdot \mathbb{E}\left[ \frac{Z_n}{p_n}\right]
.
\]
By the law of total expectation, we derive
\begin{align*}
\mathbb{E}\left[\hat{A}(t)\right]
=
\sum_{n=1,...,N} w(t, e_n) \cdot
\mathbb{E}\left[
\mathbb{E}\left[
\frac{Z_n}{p_n}\ \Big|\ \mathcal{F}_{n-1}\right]\right]
=
A(t)
.  
\end{align*}
\end{proof}

\subsection{Estimator Variance}

We show that 
\begin{equation*}
    \textup{Var}\left[ \hat{A}(t) \right] = w^2(t, e_n)
    \left(
    \mathbb{E}\left[ \frac{1}{p_n} \right] - 1
    \right).
\end{equation*}

\begin{proof}
Define
\[X_n\coloneq w(t, e_n)\frac{Z_n}{p_n},\]
s.t. $\mathbb{E}[X_n]=w(t, e_n)$. By definition: 
\begin{align*}
    \textup{Var}\left[ \hat{A}(t) \right]
	&=
	\mathbb{E}\left[
	\left(
	\hat{A}(t)-A(t)
	\right)^2
	\right]\\
	&=
    \sum_n
	\underbrace{
        	\mathbb{E}\left[\left( X_n-w(t, e_n) \right)^2\right]
    }_{\mathbb{V}(X_n)}\\
	&+
    \sum_n
    \sum_{i\neq n}
	\underbrace{
        \mathbb{E}
        \left[
		\left(
		X_n-w(t, e_n)
		\right)
		\left(
		X_i-w(t, e_i)
		\right)
        \right]
        }_{\text{Cov}(X_n,X_i)}
	.
\end{align*}
Furthermore,
\[
\begin{split}
    \mathbb{V}(X_n)
    & =
    \mathbb{E}[X_n^2]
    -
    \mathbb{E}[X_n]^2
    =
    \mathbb{E}\left[\left(w(t, e_n)\frac{Z_n}{p_n}\right)^2\right]
    -
    w^2(t, e_n)
    \\
    & =
    w^2(t, e_n)
    \left(
    \mathbb{E}\left[\frac{Z_n^2}{p_n^2}\right]
    -
    1
    \right)
    \\
    & =
    w^2(t, e_n)
    \left(
    \mathbb{E}\left[
    \mathbb{E}\left[
    \frac{Z_n^2}{p_n^2}\ \Big| \ \mathcal{F}_{n-1}
    \right]\right]
    -
    1
    \right)
    \\
    & =
    w^2(t, e_n)
    \left(
    \mathbb{E}\left[ \frac{1}{p_n} \right]
    -
    1
    \right).
\end{split}	
\]  
Additionally,
\begin{align*}
    &\mathbb{E}\left[
	\sum_n
	\sum_{i\neq n}
	\left(
	X_n-w(t, e_n)
	\right)
	\left(
	X_i-w(t, e_i)
	\right)
	\right]\\
	&=
	2
	\sum_n
	\sum_{i< n}
	\mathbb{E}\left[
	\left(
	X_n-w(t, e_n)
	\right)
	\left(
	X_i-w(t, e_i)
	\right)
	\right]
	,
\end{align*}
allowing us to focus on $\mathbb{E}\left[
\left(
X_n-w(t, e_n)
\right)
\left(
X_i-w(t, e_i)
\right)\right]$ for $i< n$ in specific.
Now:
\[
\begin{split}
    & \mathbb{E}\left[
    \left(
    X_n-w(t, e_n)
    \right)
    \left(
    X_i-w(t, e_i)
    \right)\right]\\
    &=
    \mathbb{E}\left[
    \mathbb{E}\left[
    \left(
    X_n-w(t, e_n)
    \right)
    \left(
    X_i-w(t, e_i)
    \right)
    \ \Big| \ \mathcal{F}_{n-1}
    \right]\right]
    \\
    & =
    \mathbb{E}\left[
    \left(
    X_i-w(t, e_i)
    \right)
    \mathbb{E}\left[
    \left(
    X_n-w(t, e_n)
    \right)
    \ \Big| \ \mathcal{F}_{n-1}
    \right]\right]
    \\
    & =
    0
\end{split}
\]
by noting that \(\mathbb{E}[X_n\ \Big| \ \mathcal{F}_{n-1}]=w(t, e_n)\). 
\end{proof}

\section{Kernel Density Intensity Estimators}
\label{sec:app-kde}

Again, let $\mathcal{E}$ define a Marked Inhomogeneous Counting Process
\[
    \mathcal{E} = \{ e_1, e_2, \dots \}, \quad e_n = (q_n, t_n),
\]
where $t_n \in \mathbb{R}^+$ are arrival times drawn from a time-varying intensity function $\lambda(t)$. Consider the on-line kernel density estimator:
\[
    \hat{\lambda}(t) = \frac{1}{h}\sum_{t_n < t} \exp\left(-\frac{t-t_n}{h}\right), \quad h > 0,
\]
which defines a \textbf{smoothed, past-weighted-average count} with bandwidth $h>0$. Note that, via Campbell's Theorem, it holds
\[
    \mathbb{E}[\hat{\lambda}(t)] = \frac{1}{h} \int_{-\infty}^{t} \exp\left(-\frac{t-s}{h}\right) \lambda(s)\,ds
\]
which defines a \textbf{convolution} of the true intensity $\lambda(s)$ and an exponential decay kernel. For $\hat{\lambda}(t)$ to be unbiased, the following integral equation must hold:
\[
    \lambda(t) = \frac{1}{h} \int_{-\infty}^{t} \exp\left(-\frac{t-s}{h}\right) \lambda(s)\,ds.
\]
Above, differentiation under the integral sign implies $\lambda'(t)=0$. Thus, \textit{unbiased} estimates are only retrieved in \textit{homogeneous} settings with constant rate. In time-varying, inhomogeneous settings, the convolution operation introduces a smoothing effect.

\paragraph{Unbiased Local Weighted Counts}. The estimator is widely used in practice because the term $h \cdot \hat{\lambda}(t)$ 
is a low-variance, approximately unbiased estimate of the count $N(t-h, t)$. 
Indeed,
\[
    \mathbb{E}[h \hat{\lambda}(t)]
     - \mathbb{E}[N(t-h,t)]
    = \int_{0}^{\infty} e^{-u/h}\lambda(t-u)\,du 
      - \int_{0}^{h}\lambda(t-u)\,du,
\]
which vanishes when $\lambda$ is constant. For smoothly varying $\lambda$, a Taylor expansion
$$\lambda(t-u) = \lambda(t) - u\lambda'(t) 
+ \tfrac{u^{2}}{2}\lambda''(t) + \cdots$$
yields the leading-order approximation
\[
    \mathbb{E}[h \hat{\lambda}(t)]
     - \mathbb{E}[N(t-h,t)] 
    \approx -\frac{h^{2}}{2}\,\lambda'(t).
\]
Hence the bias is $O(h^{2})$, small when $\lambda$ varies slowly relative to $h$.

Furthermore, note that
\[
\mathbb{V}\!\left(h \hat{\lambda}(t)\right)
= \int_0^\infty e^{-2u/h}\,\lambda(t-u)\,du.
\]
For constant $\lambda$, this becomes
\[
    \mathbb{V}\!\left(h \hat{\lambda}(t)\right)
    = \lambda \int_{0}^{\infty} e^{-2u/h}\,du
    = \lambda \frac{h}{2},
\]
while the variance of the true count is given by $\mathrm{Var}[N(t-h,t)] = \lambda h$. Thus, the kernel estimator \textbf{reduces the variance} by a factor of $1/2$ in the homogeneous case. Empirical evidence suggests comparable variance reduction in slowly varying inhomogeneous settings.

\section{Stability of filtered estimation}
\label{app:double-stochastic-follows-correct-counts}

We prove Remark \ref{rem:martingale} by showing that persistence-path control induces a self-correcting estimation process.

\begin{proof}
Let
\[
 \mathcal{F}^{\prime}_{n-1}
 = \sigma(e_1,\dots,e_{n-1}; Z_1,\dots,Z_{n-1})
\]
denote the natural filtration containing all information available immediately before the arrival of event $e_n$ and the corresponding filtering decision $Z_n$. Define the exponential decay factor
\[
    \beta_n=\exp\!\left(-(t_n - t_{n-1})/h\right)
\]
with KDE bandwidth parameter $h>0$. The \textit{full-stream} kernel estimator for arrival rates evolves deterministically as
\[
  \hat{\lambda}(t_n) = \frac{\hat{\nu}(t_n)}{h}, \quad
  \hat{\nu}(t_n) = 1 + \beta_n \cdot \hat{\nu}(t_{n-1}).
\]

Under persistence-path control, the filtered arrival rate estimator updates stochastically, i.e.
\[
  \hat{\lambda}_{\mathrm{F}}(t_n) = \frac{\hat{\nu}_{F}(t_n)}{h}, \quad \hat{\nu}_{\mathrm{F}}(t_n)
  = \frac{Z_n}{p_n}
    + \beta_n\cdot\hat{\nu}_{F}(t_{n-1}),
\]
where $Z_n\sim\text{Bernoulli}(p_n)$ and $p_n$ is $\mathcal{F}^{\prime}_{n-1}$-measurable.

By conditioning, and noting that 
\begin{align*}
    \mathbb{E}\left[\frac{Z_{n}}{p_{n}}\ \Big|\ \mathcal{F}^{\prime}_{n-1}\right] = 1
\end{align*}
we obtain
\begin{align}
    \mathbb{E}[\hat{\nu}_{\mathrm{F}}(t_{n}) \mid \mathcal{F}^{\prime}_{n-1}] = \beta_{n} \cdot \hat{\nu}_{\mathrm{F}}(t_{n-1}) + 1 \label{eq:mart1}.
\end{align}
We now examine the martingale property of
\[
M_n
= \frac{\hat{\lambda}_{\mathrm{F}}(t_n) - \hat{\lambda}(t_n)}
       {\exp(-t_n/h)}.
\]
Using~\eqref{eq:mart1} and the deterministic update of $\hat{\nu}(t_n)$,
\begin{align*}
    \mathbb{E}[M_n \mid \mathcal{F}^{\prime}_{n-1}]
    &= \frac{\beta_n\!\left(\hat{\nu}_{\mathrm{F}}(t_{n-1}) - \hat{\nu}(t_{n-1})\right)}
            {h\,\exp(-t_n/h)} \\
    &= \frac{\hat{\lambda}_{\mathrm{F}}(t_{n-1}) - \hat{\lambda}(t_{n-1})}
            {\exp(-t_{n-1}/h)}
    = M_{n-1}.
\end{align*}

Thus, $\{M_n\}_{n\ge0}$ is a martingale.
\end{proof}

\begin{figure}[H]
    \centering
    \includegraphics[width=.95\columnwidth]{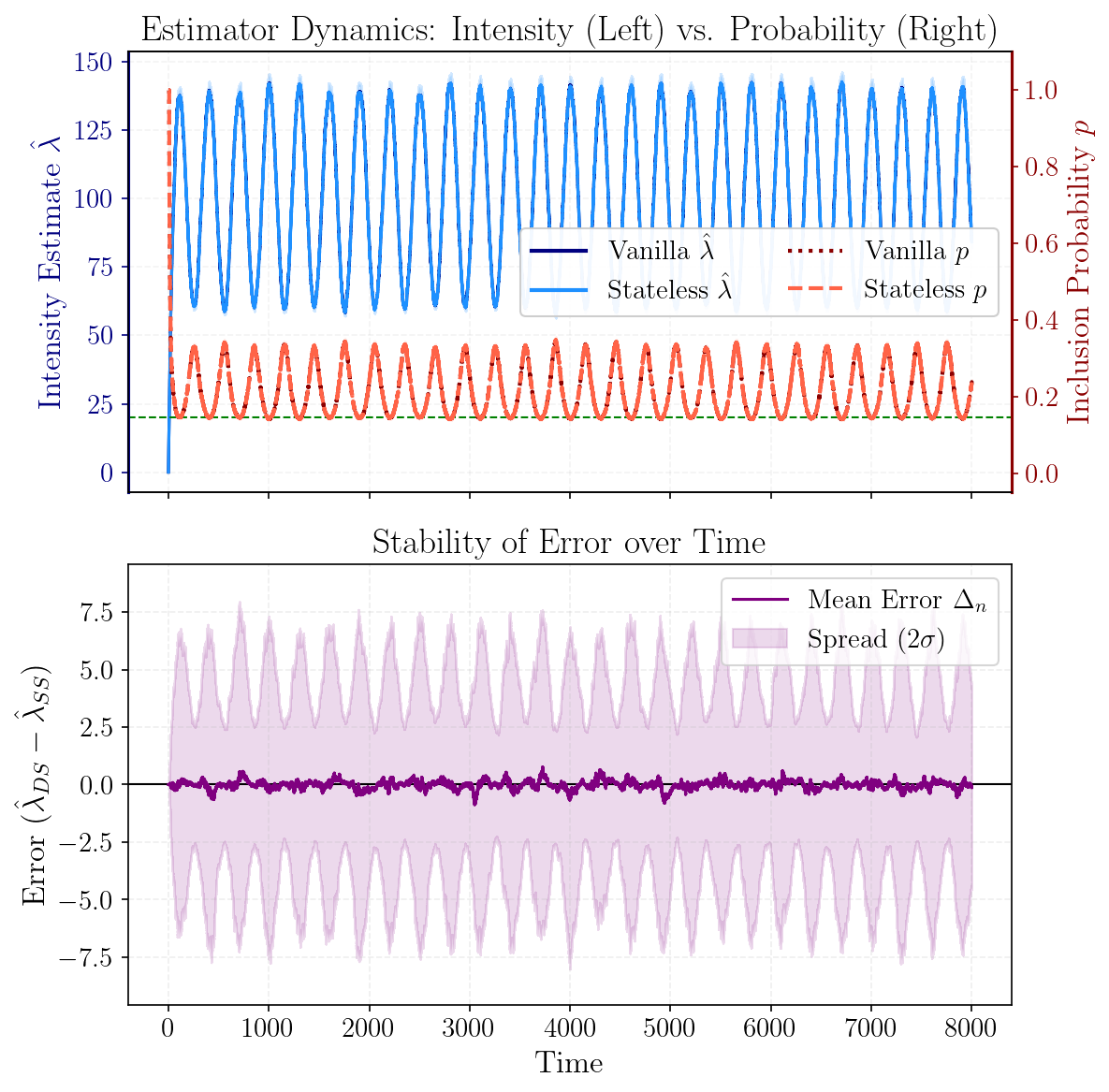}
    \caption{Top: Arrival intensity estimations under persistence-path and full-stream approaches. Bottom: Estimation difference across both approaches over a long period of time.}
    \label{fig:martingale_stability}
\end{figure}

Assuming identical initialization $\hat{\lambda}_{\mathrm{F}}(t_0)=\hat{\lambda}(t_0)$, it follows that $\mathbb{E}[M_n]=0$ for all $n$. This establishes that deviations between filtered and full-stream intensity
estimates do not compound over time, but are damped by the exponential decay of the estimator itself. In Figure~\ref{fig:martingale_stability} we illustrates the stability of this \textit{self-correcting} process over time.

\section{Safety via oversampling}
~\label{app:double-stochastic-oversampling}

We prove Remark~\ref{rem:oversampling}, showing that persistence-path control cannot reduce the expected number of persistent updates relative to full-stream control, under a naive filtering formulation.

\begin{proof}
Let $\mathcal{N}_{\mathrm{F}}$ and $\mathcal{N}$ denote the total number of updates triggered under filtered and full-stream control, respectively. Both mechanisms apply a thinning rule
\[
p_n = \min\!\left(1,\frac{\Lambda}{\hat{\lambda}(t_n)}\right),
\]
with user-defined threshold $\Lambda > 0$, but differ in the intensity estimates they rely on. A full-stream estimator $\hat{\lambda}(t_n)$ is deterministic given the arrival history, and $\hat{\lambda}_{\textbf{F}}(t_n)$ is a random estimator constructed via self-normalized filtering.

We consider the regime $\hat{\lambda}(t_n)\ge\Lambda$ almost surely; and the result trivially holds otherwise. The expected number of persistent updates under filtered control is
\[
\mathbb{E}[\mathcal{N}_{\mathrm{F}}]
= \Lambda \sum_{n} \mathbb{E}\!\left[\frac{1}{\hat{\lambda}_{\mathrm{F}}(t_n)}\right].
\]
Also, the function $g(x)=1/x$ is strictly convex for $x>0$, and by Jensen’s inequality,
\[
\mathbb{E}\!\left[\frac{1}{\hat{\lambda}_{\mathrm{F}}(t_n)}\right]
\;\ge\;
\frac{1}{\mathbb{E}[\hat{\lambda}_{\mathrm{F}}(t_n)]}.
\]

By the martingale property established in Remark~\ref{rem:martingale}, $\hat{\lambda}_{\mathrm{F}}(t_n)$ is an unbiased estimator of the full-stream intensity, i.e.
\[
\mathbb{E}[\hat{\lambda}_{\mathrm{F}}(t_n)] = \hat{\lambda}(t_n).
\]
Through substitution,
\[
\mathbb{E}[\mathcal{N}_{\mathrm{F}}]
\;\ge\;
\Lambda \sum_{n} \frac{1}{\hat{\lambda}(t_n)}
=
\mathbb{E}[\mathcal{N}].
\]

Thus, persistence-path control can only oversample relative to full-stream control in expectation.
\end{proof}

\section{Variance reduction sensitivity analysis}
~\label{app:sensitivity-analysis}

We explore the impact of the variance reduction parameter $\alpha$ in equation~($\ref{eq:variance-reduction}$) over simulated financial transaction data. We sample transaction amounts from multiple Pareto and Log-normal distributions, and simulate our persistence path mechanism with variance reduction. We compute the relative error between estimated sum features aggregates and their true unfiltered profiles. 

\begin{figure}[H]
    \centering
    \includegraphics[width=\columnwidth]{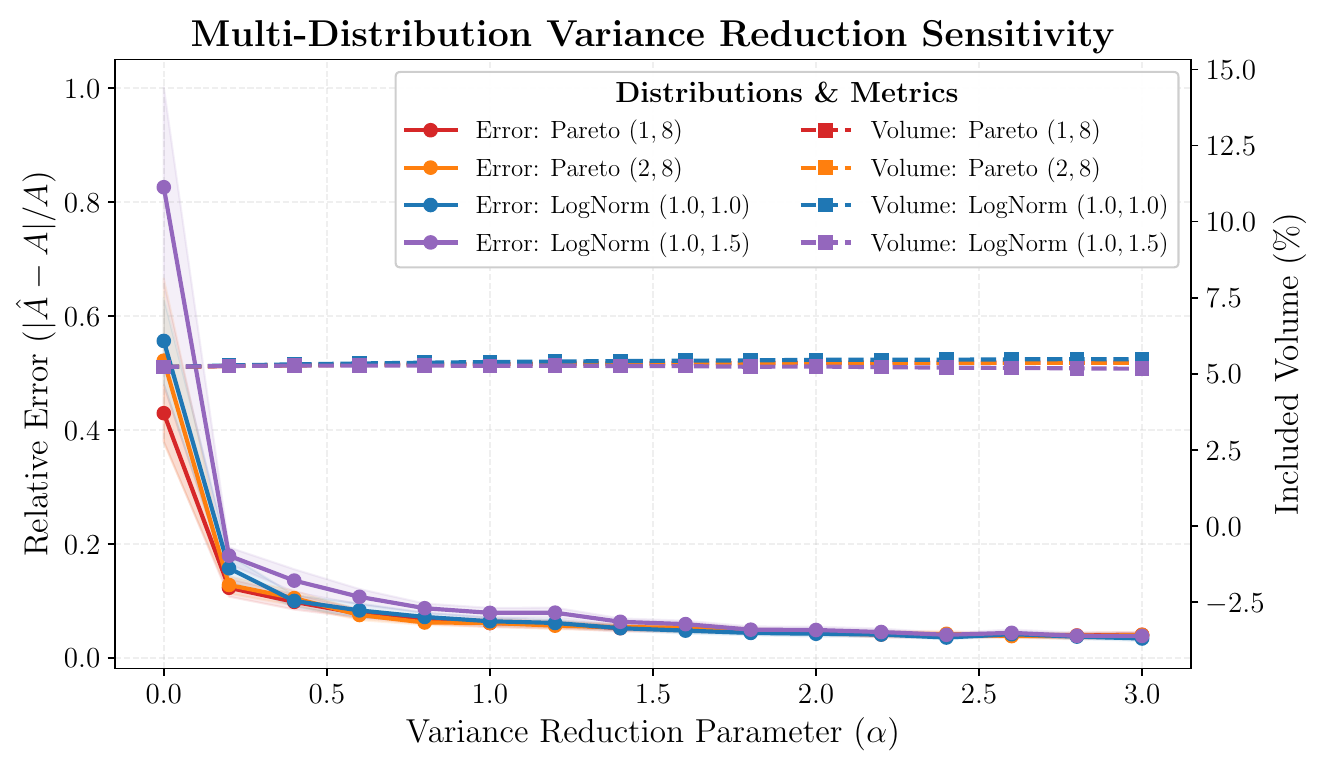}
    \caption{Estimation relative error for a decayed sum aggregation, as a function of variance-reduction parameter $\alpha$.}
    \label{fig:sens-analysis}
\end{figure}

Figure~\ref{fig:sens-analysis} shows:
\begin{itemize}
    \item The relative feature aggregation error, with $95\%$ confidence intervals, as the value of parameter $\alpha$ is increased.
    \item The volume of persistence-path operations processed, as the value of parameter $\alpha$ is increased.
\end{itemize}
The approximation error is noticeably higher for transaction amount distributions with considerable skew and kurtosis, while the number of processed persistence-path operations remains unchanged. Furthermore, increasing the effect of variance reduction yields noticeable improvements in the quality of retrieved estimates.

\begin{figure}[H]
    \centering
    \includegraphics[width=\columnwidth]{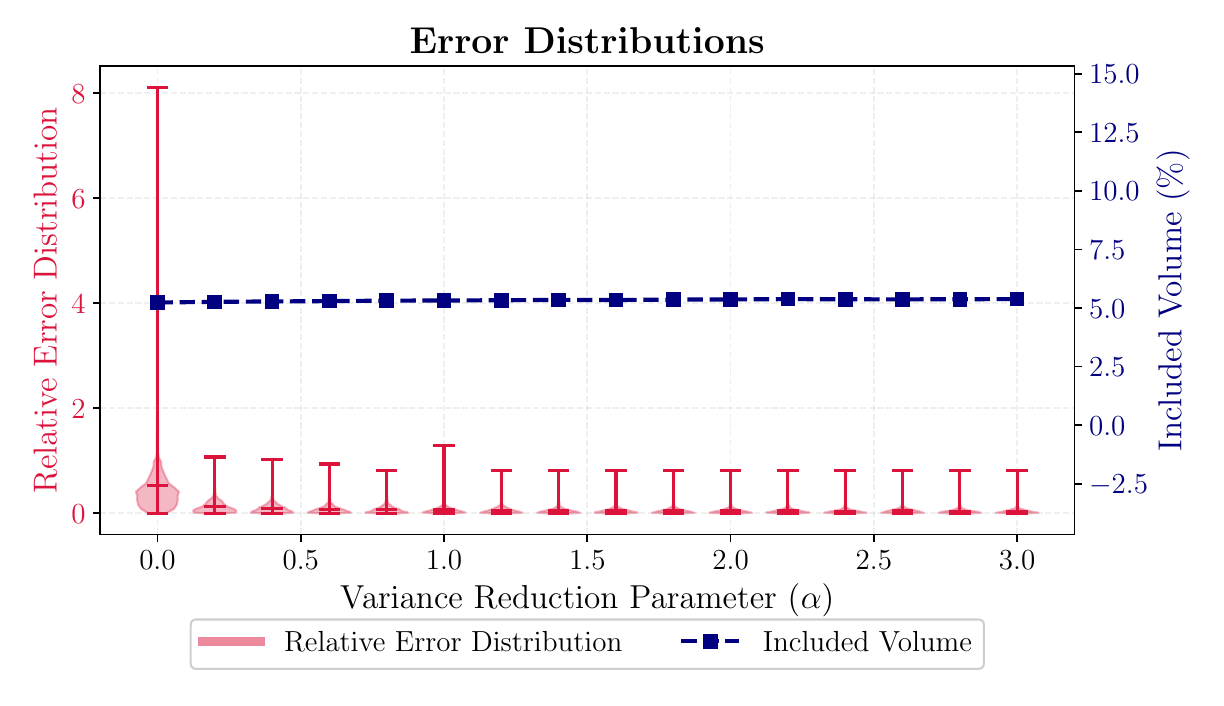}
    \caption{Error distribution for profile estimations depending on variance reduction $\alpha$ parameter.}
    \label{fig:sens-analysis-violin}
\end{figure}

Finally, Figure~\ref{fig:sens-analysis-violin} shows the relative error distribution across individual feature aggregates, for different values of $\alpha$. We notice that even midly increasing the variance reduction term can effectively ensure no extreme values are ever ignored during persistence-path aggregations; i.e. variance reduction operates by increasing the inclusion probability of outlier events.

\end{document}